\documentclass[preprint,authoryear,11pt]{elsarticle}

\usepackage{amsmath,amssymb,mathrsfs}
\usepackage{psfrag}
\usepackage{graphicx}
\usepackage{graphics}
\usepackage{epsfig}
\usepackage{bm}
\usepackage{color}
\usepackage{verbatim,color,ulem}

\newcommand{\beq}{\begin{equation}}
\newcommand{\eeq}{\end{equation}}
\newcommand{\beqa}{\begin{eqnarray}}
\newcommand{\eeqa}{\end{eqnarray}}

\journal{Physics Reports}

\begin{document}

\title{Zero-point energy of ultracold atoms}
\author{Luca Salasnich$^{1,2}$ and Flavio Toigo$^{1}$}
\address{
$^{1}$Dipartimento di Fisica e Astronomia ``Galileo Galilei''
and CNISM, \\ Universit\`a di Padova, via Marzolo 8, 35131 Padova, Italy \\
$^{2}$CNR-INO, via Nello Carrara, 1 - 50019 Sesto Fiorentino, Italy}

\date{\today}

\begin{abstract}
We analyze the divergent zero-point energy of a
dilute and ultracold gas of atoms in $D$ spatial dimensions.
For bosonic atoms we explicitly show how to regularize
this divergent contribution, which appears in the Gaussian 
fluctuations of the functional integration, 
by using three different regularization approaches:
dimensional regularization, momentum-cutoff regularization and
convergence-factor regularization.
In the case of the ideal Bose gas the divergent zero-point fluctuations
are completely removed, while in the case of the interacting Bose
gas these zero-point fluctuations give rise to a finite
correction to the equation of state. The final convergent
equation of state is independent of the
regularization procedure but depends on the dimensionality
of the system and the two-dimensional case
is highly nontrivial. 
We also discuss very recent theoretical results on the  
divergent zero-point energy of the $D$-dimensional
superfluid Fermi gas in the BCS-BEC crossover. In this case
the zero-point energy is due to both fermionic 
single-particle excitations and bosonic collective excitations,  
and its regularization gives remarkable analytical results
in the BEC regime of composite bosons. 
We compare the beyond-mean-field equations of state of both bosons 
and fermions with relevant experimental data on dilute 
and ultracold atoms quantitatively confirming the contribution of 
zero-point-energy quantum fluctuations to the thermodynamics 
of ultracold atoms at very low temperatures.
\end{abstract}


\maketitle
\tableofcontents

\section{Introduction}

The experimental achievement of Bose-Einstein
condensation \citep{bose1924,einstein1924}  with dilute and ultracold
alkali-metal atoms \citep{anderson1995,bradley1995,davis1995}
has triggered many theoretical investigations
and comprehensive reviews \citep{shi1998,dalfovo1999,leggett2001,andersen2004}
on the properties of the weakly-interacting Bose gas.
Experiments with ultracold and dilute atomic gases
in quasi-1D \citep{paredes2004,kinoshita2004}
and quasi-2D configurations \citep{hadzibabic2006,hung2011,makhalov2014}
have renewed the interest on the properties of Bose gases 
with reduced dimensionality,
where quantum fluctuations play a relevant role
\citep{mermin1966,hohenberg1967,coleman1973}.

The study of the uniform weakly-interacting Bose gas in 1, 2 and 3 dimensions  
has a long history. In their seminal papers \citet{bogoliubov1947},
\citet{ly1957} and \citet{lhy1957} investigated the properties of a
Bose gas with hard-core repulsion in three dimensions:
Bogoliubov
calculated the zero-temperature quantum depletion,
while Lee and Yang  and Lee, Huang and Yang evaluated
the leading quantum corrections to the mean-field equation by properly
treating the divergencies introduced by a naif treatment of the repulsion
as a contact interaction. In one dimension,  
based on a previous investigation of the 1D Bose-Fermi mapping by
\citet{girardeau1960},  \citet{lieb1963} obtained the exact
equation of state of a Bose gas with contact repulsive interaction 
exploiting the Bethe ansatz. In the case of two spatial dimensions,
\citet{schick1971}
found that the equation of state of a uniform 2D repulsive
Bose gas contains a nontrivial logarithmic term.
This remarkable result was improved by  \citet{popov1972} who obtained
an equation of state which, at the
leading order, reduces to  Schick's one in the dilute limit.
A key ingredient in the theoretical analysis of both \citet{schick1971}  
and \citet{popov1972} is the logarithmic
behavior of the T-matrix that describes the scattering between bosons.

In this article we show that the finite-temperature
equation of state of interacting bosons 
can be derived by functional integration \citep{nagaosa1999} of a 
$D$-dimensional model with a purely local potential after proper
regularization of the divergence of  the zero-point energy of 
Gaussian fluctuations. 
We consider three different regularization approaches: the modern
dimensional regularization \citep{thooft1972}  used mainly
in high-energy physics, the old momentum-cutoff regularization
\citep{feynman1948,pauli1949}, still
adopted in many physical contexts, and
the convergence-factor renormalization  used mainly in
condensed-matter physics \citep{nagaosa1999,stoof2009,altland2010}.
We find that the divergence
of the zero-point energy in an ideal Bose gas
is completely removed by regularization. On the contrary, in the case of
an interacting Bose gas, the zero-point energy contributes a
finite correction to the equation of state even after regularization.
The final equation of state is independent of the
regularization procedure but depends on the dimensionality
of the system. Following \citet{braaten1997}, 
\citet{andersen2004} and
\citet{schakel2008} in the one
and three-dimensional cases respectively, we recover the familiar
results of \citet{lieb1963} (in the quasi-condensate regime)
and \citet{lhy1957}, without making explicit use of
scattering theory. In the highly non-trivial two-dimensional case,
we obtain exactly Popov's equation of state \citep{popov1972} through
novel treatments of either dimensional
or cutoff regularizations. In particular, the dimensional regularization
around $D=2$ is based on a renormalization-group analysis
\citep{schakel1999,andersen2002,chien2014}
with a specific choice of the Landau pole
recently used in the study of composite bosons in the 2D BCS-BEC crossover
\citep{sala-flavio}. The momentum-cutoff  and
convergence-factor regularizations
in two dimensions are based on new approaches  developed 
for this paper. Also in the two-dimensional case, both dimensional
and momentum-cutoff regularizations do not require the use of
T-matrix scattering theory, which is instead a key ingredient
in the original derivation of  \citet{popov1972}, which is equivalent
to the convergence-factor regularization. 
Not pretending to give a complete experimental overview, 
which is far beyond the scope of this paper, 
we analyze relevant experiments with ultracold and dilute atomic 
gases in 3D (\cite{papp2008,wild2012}) and 2D (\cite{salomon2010,dalibard2011}) 
which put in evidence effects of zero-point energy on the 
equation of state of repulsive bosons. Experiments on 
1D bosons (\cite{kinoshita2004,paredes2004}) show that the Lieb-Liniger 
theory is needed to accurately describe the strong-coupling 
(i.e. low 1D density) regime. 

In this paper we also discuss current theoretical investigations
related to the regularization of the zero-point energy of a more complex
physical system: the $D$-dimensional Fermi superfluid in the BCS-BEC crossover,
i.e. the crossover of a fermionic superfluid
from weakly-bound BCS-like Cooper pairs to the Bose-Einstein condensation
(BEC) of strongly-bound molecules \citep{greiner2003,chin2004,makhalov2014}.
For this  system there are two kinds of elementary excitations
(fermionic single-particle excitations and bosonic
collective excitations) which contribute to the zero-point
energy of Gaussian quantum fluctuations.  
Very recently we have obtained remarkable results
for $D=3$ \citep{sala-giacomo} and $D=2$ \citep{sala-flavio}
removing all the divergences in the BEC regime of the crossover. 
Also for attractive fermions 
we analyze only experiments with ultracold and dilute atomic 
gases which display zero-temperature beyond-mean-field effects 
on the equation of state. The main conclusion 
is that 3D (\cite{grimm2007,leyronas2007}) and 2D 
(\cite{makhalov2014,luick2014,enss2016}) experimental data 
are in quite good agreement with the theory when Gaussian fluctuations 
are taken into account. 
For 1D superfluid fermions we show that Gaussian fluctuations
improve the mean-field theory but do not give the correct 
equation of state in the Tonks-like regime of impenetrable bosons
\citep{girardeau1960,gaudin1967}, required to reproduce 
the observed density profiles of 1D trapped atoms (\cite{liao2010}). 
General reviews of experiments with 
ultracold bosonic and fermionic atoms in reduced dimensions, achieved by using 
very anisotropic trapping potentials, can be found, for example, in 
\cite{bloch2008} and \cite{hadzibabic2011}. 

\section{Functional integration for bosonic superfluids}

We consider a D-dimensional $(D=1,2,3)$
Bose gas of ultracold and dilute
neutral atoms either noninteracting or with a repulsive
contact interaction. We adopt the path integral formalism,
where the atomic bosons are described by the complex  
field $\psi({\bf r},\tau )$ (Nagaosa, 1999).
The Euclidean Lagrangian density of the free
system in a D-dimensional box of volume $L^D$
and with chemical potential $\mu$ is given by
\beq
\mathscr{L} = \bar{\psi} \left[ \hbar \partial_{\tau}
- \frac{\hbar^2}{2m}\nabla^2 - \mu \right] \psi
+ {1\over 2} \, g \, |{\psi}|^4 \; ,
\label{lagrangian-initial}
\eeq
where $g>0$ is the strength of the contact inter-atomic
coupling (Nagaosa, 1999). The partition function ${\cal Z}$ of the
system at temperature $T$ can then be written as
\beq
{\cal Z} = \int {\cal D}[\psi,\bar{\psi}] \
\exp{\left\{ - {S[\psi, \bar{\psi}] \over \hbar} \right\}} \; ,
\label{papo}
\eeq
where
\beq
S[\psi, \bar{\psi}] =
\int_0^{\hbar\beta}
d\tau \int_{{L^D}} d^D{\bf r} \
\mathscr{L}(\psi, \bar{\psi})
\label{action}
\eeq
is the Euclidean action and
$\beta \equiv 1/(k_B T)$ with $k_B$ the Boltzmann's constant.
The grand potential $\Omega$ of the system, which is a function
of the thermodynamic variables $\mu$, $T$ and of the parameter 
$g$, is then obtained as
\beq
\Omega = -{1\over \beta} \ln{\cal Z} \; .
\eeq
We work in the superfluid phase where
the global U(1) gauge symmetry of the system is spontaneously broken
(Nagaosa, 1999). For this reason we set
\beq
\psi({\bf r},\tau) = \psi_0 +\eta({\bf r},\tau)  \; ,
\label{polar0}
\eeq
where $\eta({\bf r},\tau)$ is the complex field of bosonic
fluctuations around the order parameter $\psi_0$ (condensate in 3D or
quasi-condensate in 1D and 2D) of the system. We suppose that $\psi_0$ is
constant in time, uniform in space and real.  

\subsection{Ideal Bose gas}

First we analyze the case with $g=0$, where exact analytical
results can be obtained in any spatial dimension $D$. In fact,
the Euclidean action of the ideal Bose gas  
can be written in a diagonal form as
\beqa
S[\psi,\bar{\psi}] &=&
-\mu\, \psi_0^2 \, \hbar \beta \, L^D  
\nonumber
\\
&+& {1\over 2} \sum_{Q}
({\bar\psi}(Q),\psi(-Q)) \ {\bf M}(Q) \left(
\begin{array}{c}
\psi(Q) \\
{\bar\psi}(-Q)
\end{array}
\right) \;
\label{sigo0}
\eeqa
where $Q=({\bf q},i\omega_n)$ is
the $D+1$ vector denoting the momenta ${\bf q}$ and bosonic Matsubara
frequencies $\omega_n=2\pi n/(\beta \hbar)$, and
\beq
{\bf M}(Q) = \beta \,
\left(
\begin{matrix}
-i \hbar \omega_n + {\hbar^2q^2\over 2m} - \mu & 0 \\
0 & i \hbar \omega_n + {\hbar^2q^2\over 2m} - \mu
\end{matrix}
\right)
\eeq
is the diagonal inverse fluctuation propagator of the quadratic action.

Integrating over the bosonic fields $\eta(Q)$
and $\bar{\eta}(Q)$ in Eq. (\ref{sigo0}) one finds the grand potential
\beqa
\Omega &=& - \mu\, \psi_0^2 L^D +
{1\over 2\beta} \sum_{Q} \ln{\mbox{Det}({\bf M}(Q))}
\nonumber
\\
&=& - \mu\, \psi_0^2 L^D + 
{1\over 2\beta} \sum_{\bf q} \sum_{n=-\infty}^{+\infty}
\ln{[\beta^2 (\hbar^2\omega_n^2+\xi_{q}^2)]} \; ,
\label{pl0}
\eeqa
where $\xi_q$ is the shifted free-particle spectrum, i.e.  
\beq
\xi_{q} = {\hbar^2q^2 \over 2m} - \mu \; .
\label{spettro-bello}
\eeq
The sum over bosonic Matsubara frequencies
gives (Andersen, 2004; Kapusta, 1993; Le Bellac, 1996)
\beq
{1\over 2\beta}
\sum_{n=-\infty}^{+\infty} \ln{[\beta^2(\hbar^2\omega_n^2+\xi_{q}^2)]} =
{\xi_q\over 2} + {1\over \beta } \ln{(1-e^{-\beta\, \xi_q})} \; .
\label{flavione0}
\eeq
Strictly speaking, there should also be an additional infinite term 
on the right side of Eq. (\ref{flavione0}). However, since this infinite
constant is independent of $\beta$ and $\mu$, it can be
neglected (Kapusta, 1993; Le Bellac, 1996).
The grand potential finally reads
\beq
\Omega = \Omega_0 + \Omega^{(0)} + \Omega^{(T)} \; ,
\label{omegacol-div0}
\eeq
where
\beq
\Omega_0 = - \mu \, \psi_0^2 L^D
\eeq
is the grand potential of the order parameter,
\beq
\Omega^{(0)} = {1\over 2} \sum_{{\bf q}} \xi_q
\eeq
is the zero-point energy
of bosonic single-particle excitations, i.e. the zero-temperature
contribution of quantum fluctuations, and
\beq
\Omega^{(T)} =
{1\over \beta} \sum_{{\bf q}} \ln{\left(1 - e^{-\beta \xi_q} \right)} \;  
\label{recot}
\eeq
takes into account
thermal fluctuations. In the continuum limit, where
$\sum_{\bf q}\to L^D\int d^{D}{\bf q}/(2\pi)^D$,
the zero-point energy
\beq
{\Omega^{(0)}\over L^D} = {1\over 2} {S_D\over (2\pi)^D}
\int_0^{+\infty} dq \ q^{D-1} \left(
{\hbar^2q^2 \over 2m} - \mu \right)
\label{zz0}
\eeq
of the ideal Bose gas is clearly ultraviolet divergent
at any integer dimension $D$,
i.e for $D=1,2,3$.  Here $S_D=2\pi^{D/2}/\Gamma(D/2)$ is the solid angle
in $D$ dimensions with $\Gamma(x)$ the Euler gamma function.
We shall show that this divergent
zero-point energy of the ideal
Bose gas is completely eliminated by dimensional regularization,  
or momentum-cutoff regularization, or convergence-factor regularization.
Consequently, the exact grand potential of the ideal Bose gas is given by
\beq
{\Omega \over L^D} = - \mu \; \psi_0^2 + {1\over \beta L^D}
\sum_{{\bf q}} \ln{\left(1 - e^{-\beta \xi_q} \right)} \; .
\label{submarcos}
\eeq
We notice that $\psi_0$ is not a free parameter but must be  determined
by minimizing $\Omega_0$, namely  
\beq
\left({\partial \Omega_{0}\over \partial \psi_0}\right)_{\mu,T,L^D} = 0 \; ,  
\label{ordino-bose0}
\eeq
from which one finds that
\beq
\psi_0 =
\left\{ \begin{array}{ll}
        0 & \mbox{if $\mu <0$} \\
        \mbox{any value} & \mbox{if $\mu = 0$}  
\end{array} \right.  
\eeq

The number density $n=N/L^D$ is obtained from the thermodynamic
relation
\beq
n = - {1\over L^D} \left({\partial \Omega\over \partial \mu}\right)_{T,L^D} \; ,
\label{number-find-fla}
\eeq
which gives:
\beq
n = \psi_0^2 + {1\over L^D} \sum_{\bf q} {1\over e^{\beta \xi_q} - 1} \; .
\label{kurtona}
\eeq
We stress that, for the ideal Bose gas, only after fixing the total
density $n$ one can find the value of $\psi_0$ as a function of the
chemical potential $\mu$ and temperature $T$. Moreover, in the
continuum limit where 
$\sum_{\bf q}\to L^D\int d^{D}{\bf q}/(2\pi)^D$, by setting
$\psi_0=\mu=0$ from Eq. (\ref{kurtona}) one gets
\beq
n = \int {d^{D}{\bf q}\over (2\pi)^D} 
{1\over e^{\hbar^2 q^2\over 2m k_BT_c} - 1} \;  
\eeq
as the implicit equation determining the critical temperature $T_c$ 
for Bose-Einstein condensation. It is well known (Huang, 1987) that one finds
\beq
k_B\, T_c =
\left\{ \begin{array}{ll}
        \mbox{no solution} & \mbox{for $D=1$} \\
       \mbox{  $0$  }  & \mbox{for $D = 2$} \\
      {1\over 2\pi \zeta(3/2)^{2/3}}
{\hbar^2\over m} n^{2/3}  & \mbox{for $D=3$}  
\end{array} \right.  
\eeq
where $\zeta(x)$ is the Riemann zeta function. It is important
to stress that, also in the absence of true Bose-Einstein
condensation (as in $D=1$ and in $D=2$ for $T>0$), 
one can have quasi-condensation, i.e.
algebraic-long-range-order of the two-body density matrix, where
the use of the order parameter $\psi_0$ is still meaningful (Stoof, 2009).

\begin{figure}[t]
\centerline{\epsfig{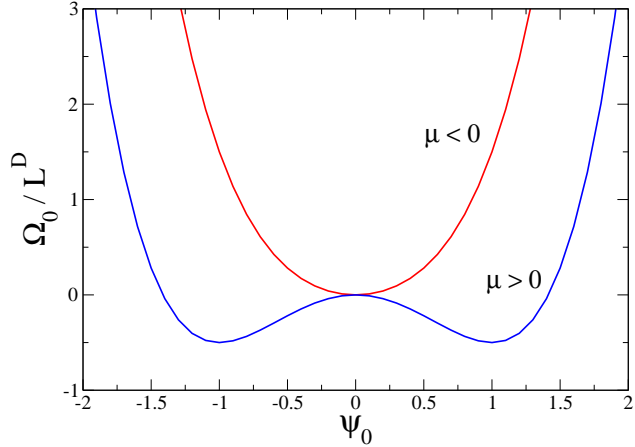}}
\small
\caption{Mean-field grand potential $\Omega_0$ as a function
of the real order parameter $\psi_0$ for an interacting Bose gas,
see Eq. (\ref{omega0}). For a positive chemical potential $\mu$
there is spontaneous symmetry breaking and the system
becomes superfluid. $\Omega$ is in units of $|\mu|$ 
and we choose $g/|\mu|=1$. For $\mu<0$ there is 
single-well potential with the minimum at $\psi_0=0$, while for 
$\mu>0$ there is a double-well potential with minima at $\psi_0\neq 0$. }
\label{fig1}
\end{figure}

\subsection{Interacting Bose gas}

Let us now consider a system of bosons with a repulsive
contact interaction, i.e. let us set  $g > 0$
in Eq. (\ref{lagrangian-initial}). 
In this case one finds immediately the partition
function of the order parameter 
\beq
{\cal Z}_{0} =  \exp{\left\{ - {S_{0}\over \hbar} \right\}}
= \exp{\left\{ - \beta \, \Omega_{0} \right\}} \; ,
\eeq
where the grand potential $\Omega_{0}$ reads (see Fig. \ref{fig1})
\beq
{\Omega_{0}\over L^D} = - \mu \, \psi_0^2 + {1\over 2} \, g \, \psi_0^4 \; .
\label{omega0}
\eeq
Again, the constant, uniform and real order
parameter $\psi_0$  is obtained by minimizing $\Omega_{0}$ as
\beq
\left({\partial \Omega_{0}\over \partial \psi_0}\right)_{\mu,T,L^D} = 0 \; ,  
\label{ordino-bose}
\eeq
from which one finds the relation between order parameter and
chemical potential
\beq
\mu = g \, \psi_0^2 \; 
\label{0o0}
\eeq
showing that in the superfluid broken phase the chemical potential
is positive and
\beq
\psi_0 = \sqrt{\mu \over g} \; . 
\label{ordereq}
\eeq
Inserting this relation into Eq. (\ref{omega0}) we find
\beq
{\Omega_{0}\over L^D} = - {\mu^2 \over 2\, g} \; .
\label{omega-mf}
\eeq
Clearly, this equation of state
is lacking important informations
encoded in quantum and thermal fluctuations.

As previously pointed out, the main goal of this paper
is to discuss and take into account these fluctuations, and in particular
zero-temperature quantum fluctuations which are
crucial in reduced dimensionalities
(Mermin and Wagner, 1966; Hohenberg, 1967; Coleman, 1973).
To this end we allow $\eta({\bf r},\tau)\neq 0$ in Eq. (\ref{polar0}) 
and expand the action $S[\psi, \bar{\psi}]$
of Eq. (\ref{action}) around $\psi_0$ 
up to  quadratic (Gaussian) order in $\eta({\bf r},\tau)$
and $\bar{\eta}({\bf r},\tau)$. One finds
\beq
Z = Z_{0} \ \int
{\cal D}[\eta,\bar{\eta}] \
\exp{\left\{ - {S_g[\eta,\bar{\eta}] \over \hbar} \right\}} \; ,
\label{sigo}
\eeq
where
\beq
S_{g}[\eta,\bar{\eta}] = {1\over 2} \sum_{Q}
({\bar\eta}(Q),\eta(-Q)) \ {\bf M}(Q) \left(
\begin{array}{c}
\eta(Q) \\
{\bar\eta}(-Q)
\end{array}
\right) \;
\eeq
is the Gaussian action of fluctuations in reciprocal space
with $Q=({\bf q},i\omega_n)$
a $D+1$ vector denoting momenta ${\bf q}$ and bosonic Matsubara
frequencies $\omega_n=2\pi n/(\beta \hbar)$, and again 
\beq
{\bf M}(Q) = \beta \,
\left(
\begin{matrix}
-i \hbar \omega_n + {\hbar^2q^2\over 2m} - \mu + 2g\psi_0^2& g\psi_0^2 \\
g\psi_0^2 & i \hbar \omega_n + {\hbar^2q^2\over 2m} - \mu + 2g\psi_0^2  
\end{matrix}
\right)
\eeq
is the inverse fluctuation propagator.

Integrating over the bosonic fields $\eta(Q)$
and $\bar{\eta}(Q)$ in Eq. (\ref{sigo}) one finds
the Gaussian grand potential
\beqa
\Omega_g &=& {1\over 2\beta} \sum_{Q} \ln{\mbox{Det}({\bf M}(Q))}
\nonumber
\\
&=& {1\over 2\beta} \sum_{\bf q} \sum_{n=-\infty}^{+\infty}
\ln{[\beta^2 (\hbar^2\omega_n^2+E_{q}^2)]} \; ,
\label{pl}
\eeqa
where 
\beq
E_{q} =
\sqrt{\left( {\hbar^2q^2 \over 2m} -\mu
+ 2 g \psi_0^2\right)^2 - g^2 \psi_0^4} \; 
\label{spettrob0}
\eeq
is the familiar Bogoliubov spectrum when Eq. (\ref{0o0}) is used. 
Taking into account Eq. (\ref{flavione0}), the sum over bosonic Matsubara
frequencies gives (Andersen, 2004; Kapusta, 1993; Le Bellac, 1996)
\beq
{1\over 2\beta}
\sum_{n=-\infty}^{+\infty} \ln{[\beta^2(\hbar^2\omega_n^2+E_{q}^2)]} =
{E_q\over 2} + {1\over \beta } \ln{(1-e^{-\beta E_q})} \; ,  
\label{flavione}
\eeq
and the total grand potential may then be written as
\beq
\Omega = \Omega_{0} + \Omega_{g}^{(0)} + \Omega_{g}^{(T)} \; ,
\label{omegacol-div}
\eeq
where $\Omega_{0}$ is given by Eq. (\ref{omega-mf}). 
\beq
\Omega_{g}^{(0)} = {1\over 2} \sum_{{\bf q}} E_q
\eeq
again is the zero-point energy 
of bosonic collective excitations, i.e. the zero-temperature contribution of
quantum Gaussian fluctuations, while
\beq
\Omega_{g}^{(T)} =
{1\over \beta} \sum_{{\bf q}} \ln{\left(1 - e^{-\beta E_q} \right)} \;  
\eeq
takes into account
thermal Gaussian fluctuations. Note that if $g=0$
one finds $E_q=\xi_q$ and Eq. (\ref{recot}) is recovered.  

\begin{figure}[t]
\centerline{\epsfig{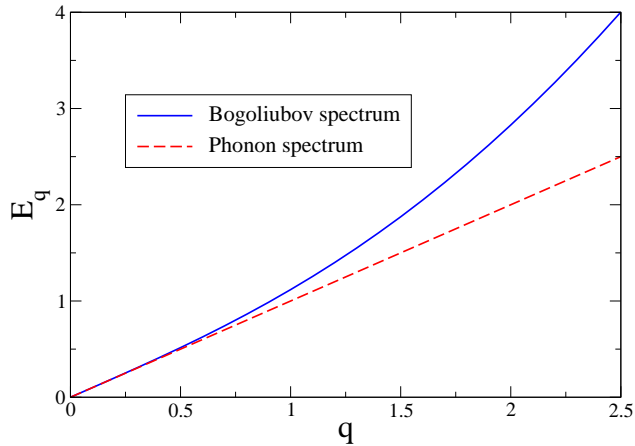}}
\small
\caption{Bogoliubov spectrum, given by Eq. (\ref{spettrob}), and
its low-momentum phonon spectrum $E_q = c_B \, \hbar q$,
where $c_B=\sqrt{\mu/m}$ is the sound velocity. 
Energy $E_q$ in units of $\mu$ and momentum $q$ in units of 
$\sqrt{m \mu/\hbar^2}$.}
\label{fig2}
\end{figure}

We remark again that using Eq. (\ref{0o0}) to remove
the dependence on $\psi_0$ in $E_q$ one obtains the usual
form for the Bogoliubov spectrum, i.e.
\beq
E_{q} =
\sqrt{{\hbar^2q^2 \over 2m}
\left( {\hbar^2q^2 \over 2m} + 2 \mu \right)} \; .
\label{spettrob}
\eeq

We notice that the continuum limit  of  
the zero-point energy for the interacting Bose gas
\beq
{\Omega_{g}^{(0)}\over L^D} = {1\over 2} {S_D\over (2\pi)^D}
\int_0^{+\infty} dq \ q^{D-1}
\sqrt{{\hbar^2q^2 \over 2m} \left( {\hbar^2q^2 \over 2m} + 2 \mu \right)}
\label{zz}
\eeq
is ultraviolet divergent at any integer dimension $D$.

We stress the formal similarities
between Eq. (\ref{zz0}) for the ideal Bose gas and Eq. (\ref{zz}) for the
interacting Bose gas. In Sections
3 and 4 we shall show that the divergent zero-point energy of
the interacting Bose gas, Eq. (\ref{zz}), gives rise to a finite
correction to the equation of state, which is independent of the
regularization procedure.

Before concluding this section we wish to comment on two points: 
a) the limits of validity of the Gaussian approximation, 
i.e. of Bogoliubov approximation, and b) the relation between 
Bogoliubov's spectrum and the speed of sound.

Regarding point a), we observe that the Gaussian approximation 
may be seen as an expansion in terms of the adimensional parameter 
$\gamma=(m/{\hbar^2}) g n^{(D-2)/D}$ with $n$ the number density of bosons. 
So, we expect Bogoliubov's approximation 
to be asymptotically correct when $ \gamma \ll 1$.

As for point b) we notice that in the low-momenta regime, 
when $\hbar q \ll \sqrt{2m \mu}$, Bogoliubov's spectrum of 
Eq. (\ref{zz}) reduces to the familiar linear phonon spectrum 
$E_q= \hbar \, c_B q$ (see Fig. \ref{fig2}).  
We stress that in general $c_B$ differs from the speed of 
sound $c_s$  derived from thermodynamics according to 
\beq
mc_s^2= 
\frac{\left({\partial \Omega\over \partial \mu} \right)_{T,L^D}}
{\left({\partial^2 \Omega\over \partial \mu^2} \right)_{T,L^D}} \, .
\label{cs}
\eeq
Indeed, even at $T=0$, $c_s$ is a complicate function of $\mu$ 
and therefore of the density $n$, since the derivatives of the grand 
potential have contributions both from $\Omega_0$ and from 
$\Omega _g^{(0)}$. Only for densities such that the latter 
is negligible  one finds $c_s$=$c_B$.  
 
\section{Regularization}

The choice of a contact interaction has allowed us to perform analytical
derivations, at the price, however of obtaining divergent results.

To get finite physical values, we must therefore regularize the
otherwise divergent Eq. (\ref{zz0}) for the ideal Bose gas and Eq. (\ref{zz})
for the interacting Bose gas by properly renormalizing the coupling
constant $g$ on the basis of some physical constant characterizing
the two body scattering, such as the scattering length.

\subsection{Regularization and scattering theory}

We recall that in 3D \citet{bogoliubov1947}  was 
able to remove divergent zero-point fluctuations by using canonical
quantization and a regularization procedure relating the physical 
s-wave scattering length $a_B$ of the actual interatomic potential 
to the strength $g$ of the model contact interaction: 
\beq
{m\over 4\pi \hbar^2 a_B}
= {1\over g} + {1\over L^3} \sum_{\bf q} {m\over \hbar^2 q^2} \;  . 
\label{again}
\eeq
This equation may be easily deduced from the Lippman-Schwinger equation 
for the $T$-matrix ${\hat T}(E)$ (see \cite{stoof2009}). 
In the low energy limit, the Lippman-Schwinger equation  
renormalizes the coupling constant of a 
contact interaction $V({\bf r})=g \ \delta({\bf r})$ as 
\beq
\frac{1}{g_r(E)}\equiv \frac{1}{T(E + i 0^+)}= {1\over g} + 
{1\over L^D} \sum_{\bf q} {m\over \hbar^2 q^2 - E - i 0^+ } \; ,
\label{again0}
\eeq
where $T(E)=T_{{\bf k},{\bf k}'}(E) \equiv \langle {\bf k}|
{\hat T(E)}|{\bf k}'\rangle $  when $k=k'$ and $E=\hbar^2k^2/m$. 

At very low energy one may relate the s-wave scattering length $a_B$, 
a physical measurable quantity, to $T(E)$ according to 
\beq 
T(E+i0^+) = 
\left\{ \begin{array}{ll}
{4\pi\hbar^2\over m} {a_B\over 1 + i a_B \sqrt{m E\over \hbar^2}}
& \mbox{ for  $D=3$} \\
-{2\pi\hbar^2\over m} 
{1\over \ln{(a_B \sqrt{m E\over \hbar^2} e^{\gamma}/2 )} + i \pi/2 }  
& \mbox{for $D=2$} \\
-{2\hbar^2\over m} {i \sqrt{m E\over \hbar^2} \over 
1 +i a_B \sqrt{mE\over \hbar^2}} 
& \mbox{ for  $D=1$}  
\end{array} \right.
\label{tmatrix}  
\eeq
with $\gamma=0.577...$ the Euler-Mascheroni constant 
(\cite{mora2003,girardeau2004,pricoupenko2011}). 

In 3D one can safely set $E=0$, 
so that $T(0+i0^+)={4\pi \hbar^2 a_B / m}$, and 
Eq. (\ref{again}) is immediately recovered. In 2D and 1D one must use 
Eq. (\ref{again0}) with a finite value of $E$, while in 1D one gets 
$T(E+i0^+)=-2\hbar^2/(ma_B)$ under the condition 
$a_B \sqrt{mE/\hbar^2} \gg 1$. 

The second term in Eq. (\ref{again}) is UV divergent, 
and this divergence 
may be regularized by introducing a hard cutoff $\Lambda$ for
momenta (Momentum Cutoff Regularization, MCR). In this way one
renormalizes $g=g(\Lambda)$ to reproduce the physical scattering length
$a_B$ and lets $\Lambda$ go to infinity at the end of the calculation. 
This method will be discussed in detail in Subsection 3.3.

In Subsection 3.4 we shall see how one can
recover Bogoliubov's result by
introducing convergence factors into Eq. (\ref{pl}) and then using
Eq. (\ref{again}) to fully remove the divergence of quantum fluctuations
(Convergence factor regularization, CFR).

Before discussing in details these methods, we illustrate a
regularization procedure largely employed in the renormalization program
of gauge theories: the method of Analytic or Dimensional
Regularization (DR). We refer to Leibbrandt (1975) for a review.  

\subsection{Dimensional regularization (DR)}

Analytic regularization exploits the concept of analytic continuation
in some complex parameter. In its best known version, developed by
't Hooft and Veltman (1972), the parameter is the space (or space-time in
the relativistic case) dimension $D$ and for this reason this procedure
is also known as dimensional regularization.

In this section we illustrate the main ingredients of this method
to get finite physical values from the otherwise divergent
Eq. (\ref{zz0}) for the ideal Bose gas and Eq. (\ref{zz})
for the interacting Bose gas.

We begin by recalling the fundamental 't Hooft and Veltman (1972)
conjecture stating that in the context of dimensional 
regularization integrals over a polynomial identically vanish, i.e.
\beq
\int_0^{+\infty} dq \, q^{D-1} \, (q^2)^{n-1} = 0, \qquad\quad n=0,1,2...
\label{dreg1}
\eeq
where $D$ may assume non integer values. The proof of this 
astonishing relation can be found in Sect IV of \citet{leibbrandt1975})). 
As a first consequence of Eq. (\ref{dreg1}) one gets from 
Eq. (\ref{again}) in $D=3$
\beq
g={ 4\pi \hbar^2 a_B \over m}    
\label{again4}
\eeq
implying that in dimensional regularization the strength $g$  of a 3D  
contact interaction is  the bare scattering length $a_B$, apart from a
multiplicative constant inversely proportional to the mass of the
interacting particles.

As a second ingredient of dimensional regularization, one notices that
for any   complex $z$ with $Re(z)>0$ one may write 
\beq
a^{-z}=\frac{1}{\Gamma(z)}\int_0^{+\infty} t^{z-1} e^{-at}
\label{rep}
\eeq
where
\beq
\Gamma(z)= \int_0^{+\infty} t^{z-1} e^{-z}, \qquad  Re(z)>0
\label{Egam}
\eeq
is Euler's Gamma function.
To discuss situations where $Re(z) \leq 0$, we must analytically continue
the definition of Eq.(\ref{Egam})
to the left part of the $z$ plane. Such continuation is found to be
analytic everywhere except at the points where $z$ is a negative integer
or zero. We will then treat such points with particular care.

Using Eq. (\ref{rep}) one gets immediately that Euler's beta function:
\beq
B(x,y) = \int_0^{+\infty} dt {t^{x-1}\over (1+t)^{x+y}}, 
\qquad Re(x), Re(y) >0
\label{Ebeta}
\eeq
may be continued to complex values of $x$ and $y$ as (Ryder, 2001)
\beq
B(x,y) = {\Gamma(x) \, \Gamma(y)\over \Gamma(x+y)}
\label{god}
\eeq

Using Eq. (\ref{Ebeta}) we may rewrite  Eq. (\ref{zz}) for the zero
point energy of a repulsive Bose gas in dimension $D$ as:  
\beq
{\Omega_{g}^{(0)}\over L^D} =
{S_D (2\mu)^{{D\over 2}+1}\over 4(2\pi)^D}
\left({2m\over \hbar^2}\right)^{{D\over 2}}
B\Big({D+1\over 2},-{D+2\over 2}\Big) \; ,
\label{ffp}
\eeq

Let us now analyze in detail the consequences of Eqs. (\ref{dreg1})
and (\ref{ffp}).

\subsubsection{Ideal Bose gas - DR}

The zero point energy Eq. (\ref{zz0}) of the Ideal Bose gas is an integral
over a sum of poynomials in $q^2$. From 't Hooft and Veldman conjecture,
Eq. (\ref{dreg1}),
it follows that in dimensional regularization its value is zero in any
of the spatial dimensions $D=1, 2, 3$
\beq
{\Omega^{(0)}\over L^D} = 0 \; . 
\label{conjecture}
\eeq 
As said before, the
zero-point fluctuations of the ideal $D$-dimensional Bose gas are then
fully removed by regularization
and the exact grand potential is given by Eq. (\ref{submarcos}).  

\subsubsection{Repulsive Bose gas in 1D and 3D - DR}

In the case of an interacting Bose gas, we rewrite Eq. (\ref{zz}) using
Eq. (\ref{god}) to get
\beq
{\Omega_{g}^{(0)}\over L^D} =
{S_D(2\mu)^{{D\over 2}+1}\over 4(2\pi)^D}
\left({2m\over \hbar^2}\right)^{{D\over 2}}
{\Gamma({D+1\over 2}) \, \Gamma(-{D+2\over 2})\over \Gamma(-{1\over 2})} \; .
\label{ff}
\eeq
This expression is now finite when $D=1$ or $D=3$,
while it is still divergent if $D=2$ since $\Gamma (p)$ diverges
for integers $p \leq 0$.

For the repulsive Bose gas in $D=1$,
from Eq. (\ref{ff}) we immediately find  
\beq
{\Omega_g^{(0)} \over L} = - {2\over 3\pi} ({m\over \hbar^2})^{{1\over 2}}
\mu^{3/2} \; .
\label{pinco}
\eeq
The corresponding total grand potential thus reads
\beq
{\Omega \over L} = - {\mu^2\over 2\, g}
- {2\over 3\pi} ({m\over \hbar^2})^{{1\over 2}} \mu^{3/2}
+ {1\over \beta L} \sum_{{\bf q}} \ln{\left(1 - e^{-\beta E_q} \right)} \; .
\label{regdim-1d}
\eeq
One may show that in the range of validity of Bogoliubov approximation, 
i.e. $\gamma(D=1) = \frac{m}{\hbar^2} \frac{g}{n} \ll 1$ with $n$ 
the 1D number density, this expression agrees with 
the exact result obtained by Lieb and Liniger (1963). 
On the contrary, as a consequence of the failure of the Gaussian 
approximation, in the limit of large $\gamma$, i.e. small $n$, one does 
not recover the correct grand potential of the Tonks-Girardeau gas 
(\cite{girardeau1960}) at zero temperature, but obtains 
instead twice its value. 

In the three-dimensional case, setting $D=3$ in Eq. (\ref{ff}) we get
instead
\beq
{\Omega_g^{(0)} \over L^3} = {8\over 15 \pi^2}
({m\over \hbar^2})^{3/2} \mu^{5/2} \; .
\label{pinc3}
\eeq
Notice that Gaussian quantum fluctuations, Eqs. (\ref{pinco}) and 
(\ref{pinc3}), contribute with
different signs to the grand-potentials in 1D and 3D. 
The total grand potential of the three-dimensional Bose
gas is then given by
\beq
{\Omega \over L^3} = - {\mu^2\over 2\, g}
+ {8\over 15 \pi^2} ({m\over \hbar^2})^{3/2} \mu^{5/2}
+ {1\over \beta L^3} \sum_{{\bf q}} \ln{\left(1 - e^{-\beta E_q} \right)} \; .
\label{regdim-3d}
\eeq
This is exactly the grand potential obtained by \citet{lhy1957}
by considering the contributions of the zero point energy of Bogoliubov
(Bogoliubov, 1947) excitations and, contrary to the 1D case, 
it is reliable in the low-density regime (i.e. for 
$\gamma(D=3)=\frac{m}{\hbar^2} g n^{\frac{1}{3}}\ll 1$). 

\subsubsection{Repulsive Bose gas in 2D - DR}

Dimensional regularization  of the repulsive Bose gas
is more delicate in 2D. In fact, for $D=2$
Eq. (\ref{ff})  diverges due to the presence of
$\Gamma(-2)$. To cure this divergence, one then extends the calculation
to non-integer dimension $D=2-\varepsilon$
and lets $ \varepsilon$ go to zero at the end of the calculation.
Eq. (\ref{ff}) can be written as
\beq
{\Omega_{g}^{(0)} \over L^D} =
- {m \over 4\pi\hbar^2 \, \kappa^{\varepsilon} } \mu^2 \
\Gamma(-2+{\varepsilon\over 2} ) \; ,
\label{om_eps}
\eeq
where the regulator $\kappa$ is an arbitrary scale wavenumber
which enters for dimensional reasons.
Since in the limit $\varepsilon \to 0$ one has:
\beq
\Gamma(-2+ {\varepsilon \over 2})= {1\over \varepsilon}
+ O(\varepsilon^0),
\eeq
to leading order in $1/\varepsilon$ we get
\beq
{\Omega_{g} \over L^D} = - {1\over 2} {m\over 2\pi\hbar^2 \varepsilon \,
\kappa^{\epsilon}} \, \mu^2 \; .
\label{biro}
\eeq
Comparing $\Omega_g^{(0)}$ with $\Omega_{0}$ in $D$ dimensions
we conclude that
\beq
{\Omega_{0}\over L^2} +
{\Omega_g \over L^2} = - {1\over 2 g_r} \, \mu^2 \; ,
\label{totale}
\eeq
where the renormalized coupling
constant $g_r$ given by
\beq
{1\over g_r}
= \kappa^{\epsilon} \left(
{1\over g} + {m \over 2\pi \hbar^2 \kappa^{\varepsilon} \,
\varepsilon} \right) \; ,  
\label{rodi}
\eeq
The parameter $g_r$ is a ``running coupling constant'' for our
theory which varies by changing $\kappa$.
To extract its dependence on $\kappa$ within a renormalization-group
scheme we obtain from (\ref{rodi}) the differential flow equation
\beq
\kappa {dg_r\over d\kappa} = {m\over 2\pi\hbar^2} \ g_r^2.
\eeq
In the limit $\varepsilon \to 0$ we get the
solution
\beq
{1\over g_r(\kappa_0)} - {1\over g_r(\kappa)}
= - {m\over 2\pi\hbar^2} \ \ln{\left({\kappa_0\over \kappa}\right)} \; .  
\label{simsalabim}
\eeq
By setting the Landau pole  (Kaku, 1993)
of Eq. (\ref{simsalabim}) at the energy $\epsilon_0= \hbar^2{\kappa_0}^2/(2m)$, 
i.e. by defining
\beq 
{1\over g_r(\kappa_0)} = 0 \; , 
\eeq
we obtain
\beq
{1\over g_r(\kappa)} =  
{m\over 4\pi\hbar^2} \ \ln{\left({\epsilon_0\over \mu}\right)} \; 
\label{ziobilly}
\eeq
when $\kappa$ is such that $\hbar^2\kappa^2/(2m)=\mu$. 
Thus, the running coupling constant $g_r$ is indeed a function
of the chemical potential $\mu$ and of an energy  $\epsilon_0$.

By inserting the renormalized coupling constant $g_r(\kappa)$ from
Eq. (\ref{ziobilly}) into
Eqs. (\ref{omegacol-div}) and (\ref{totale})
we thus obtain the regularized
beyond-mean-field grand potential in the form
\beq
{\Omega\over L^2} = - {m \over 8\pi\hbar^2}
\ln{\left({\epsilon_0\over \mu } \right)}\, \mu^2 +
{1\over \beta L^2} \sum_{{\bf q}} \ln{\left(1 - e^{-\beta E_q} \right)} \; .
\label{cicredi}
\eeq
This is exactly the equation of state derived by Popov (Popov, 1972)
from a 2D hydrodynamic Hamiltonian taking into account
the logarithmic behavior of the T-matrix that describes
the scattering of two bosons (Schick, 1971). The energy
$\epsilon_0$ appearing
in Eq. (\ref{cicredi})  was introduced by Popov (Popov, 1972) as a
 cutoff in the T-matrix. By equating $g_r(\kappa)$ above with 
the 2D expression of $T(\mu)$ from Eq. (\ref{tmatrix}) 
we identify $\epsilon_0$ as: $\epsilon_0=\hbar^2/(m {a_B}^2 e^{\gamma})$.
 
Notice that any dependence on the bare interaction
strength $g$ has completely disappeared from the final expression
Eq. (\ref{cicredi}) of the grand potential.

\begin{table}[h]
\begin{center}
\begin{tabular}{|c|c|}
\hline\hline
Dimension & Grand potential \\
\hline
$D=3$ & ${\Omega\over L^3}= - {\mu^2\over 2\, g}
+ {8\over 15 \pi^2} ({m\over \hbar^2})^{3/2} \mu^{5/2}
+ {1\over \beta L^3} \sum_{{\bf q}} \ln{\left(1 - e^{-\beta E_q} \right)}$
\\
\hline
$D=2$ &
${\Omega\over L^2} = -{m \over 8\pi\hbar^2}
\ln{\left({\epsilon_0\over \mu } \right)}\, \mu^2 +
{1\over \beta L^2} \sum_{{\bf q}} \ln{\left(1 - e^{-\beta E_q} \right)}$
\\
\hline
$D=1$ &
${\Omega\over L}=- {\mu^2\over 2\, g}
- {2\over 3\pi} ({m\over \hbar^2})^{{1\over 2}} \mu^{3/2}
+ {1\over \beta L} \sum_{{\bf q}} \ln{\left(1 - e^{-\beta E_q} \right)}$
\\
\hline\hline
\end{tabular}
\end{center}
\caption{Grand potential $\Omega$ of the $D$-dimensional interacting Bose
gas, with Bogoliubov spectrum
$E_q=\sqrt{{\hbar^2k^2\over 2m}({\hbar^2k^2\over 2m}+2\mu)}$,
obtained after regularization of zero-point Gaussian fluctuations.
$\mu$ is the chemical potential and $\beta=1/(k_BT)$ with $k_B$ the
Boltzmann constant and $T$ the absolute temperature.}
\end{table}

For the sake of completeness, in Table 1 we report the final equation
of state for $D=1,2,3$ of the interacting Bose gas at the gaussian level.

\subsection{Momentum-cutoff regularization (MCR)}

In this subsection we show how to regularize the divergent zero-point
energy of the bosonic gas by means of an ultraviolet cutoff $\Lambda$
in the wavenumber $q$
and a subsequent renormalization of the bare parameters of the theory.

\subsubsection{Ideal Bose gas - MCR}

By using a high wavenumber cutoff $\Lambda$, the zero-temperature
contribution of quantum fluctuations to the grand potential  of the
ideal Bose gas reported in Eq. (\ref{zz0}) becomes
\beqa
{\Omega^{(0)}\over L^D}= {1\over 2} {S_D\over (2\pi)^D}
\Big( {\hbar^2\over 2m} \int_0^{\Lambda} dq \ q^{D+1}
-\mu \int_0^{\Lambda} dq \ q^{D-1} \Big)
\nonumber
\\
= {1\over 2} {S_D\over (2\pi)^D}
\left( {\hbar^2\over 2m (D+2)} \Lambda^{D+2} - \mu
\int_0^{\Lambda} dq \ q^{D-1}
\right) \;.
\label{succhia}
\eeqa
The term proportional to $\Lambda^{D+2}$ is independent of $\mu$
and can therefore be ignored. The other term depends linearly on $\mu$
but it can be absorbed into $\Omega^{(0)} \over L^D$ by redefining the
order parameter $\psi_0$ in the zero-temperature grand potential
\beq
{\Omega_0\over L^D} + {\Omega^{(0)} \over L^D}
= - \mu \, \psi_0^2 - \mu {1\over 2} {S_D\over (2\pi)^D}
\int_0^{\Lambda} dq \ q^{D-1} \; .
\eeq
In fact, by setting
\beq
\psi_0^2 = \psi_{0,r}^2 + \delta \psi_0^2
\eeq
with the counterterm
\beq
\delta \psi_0^2 = - {1\over 2} {S_D\over (2\pi)^D}
\int_0^{\Lambda} dq \ q^{D-1}
\eeq
one finds the total grand potential in the form
\beq  
{\Omega \over L^D} = - \mu \; \psi_{0,r}^2 + {1\over \beta L^D}
\sum_{{\bf q}} \ln{\left(1 - e^{-\beta \xi_q} \right)} \; .  
\eeq
This is exactly Eq. (\ref{submarcos}), apart for the subscript
$r$ in $\Omega$ and $\psi_0$. In Section II
we have found the same grand potential by using dimensional regularization.

\subsubsection{Repulsive Bose gas in 1D and 3D - MCR}

For the interacting Bose gas Eq. (\ref{zz}) gives
\beqa
{\Omega_{g}^{(0)}\over L^D} &=& {1\over 2} {S_D\over (2\pi)^D}
\int_0^{\Lambda} dq \ q^{D-1}
\sqrt{{\hbar^2q^2 \over 2m} \left( {\hbar^2q^2 \over 2m} + 2 \mu \right)}
\nonumber
\\
&=& {S_D (2\mu)^{{D\over 2}+1}\over 4(2\pi)^D}
\left({2m\over \hbar^2}\right)^{{D\over 2}}
\int_0^{Z} {\hskip -0.2cm} dz \, z^{D-1\over 2} \sqrt{1+z} \; ,
\label{succhia1}
\eeqa
where $Z=\hbar^2\Lambda^2/(4 m \mu)$.
These expressions are now well defined for $D=1,2,3$ but obviously
diverge as $Z\to +\infty$.  As we shall see however, also in this case
the divergent contributions may be removed by introducing appropriate
counterterms (Feynman, 1948; Kaku, 1993; Schakel, 2008) which renormalize
the bare parameters of the theory.

For the one-dimensional repulsive Bose gas
we set $D=1$ in Eq. (\ref{succhia1}) and after integration we find  
\beq
{\Omega_g^{(0)}\over L} = {2\over 3\pi} ({m\over \hbar^2})^{1/2} \mu^{3/2}
\left(-1 + \Big(1+ Z \Big)^{3/2} \right) \; .  
\eeq
In the large-$\Lambda$ limit, the zero-temperature total grand potential
is then given by
\beqa
{\Omega_{0}\over L} + {\Omega_g^{(0)}\over L} &=& -{\mu^2\over 2\, g}
- {2\over 3\pi} ({m\over \hbar^2})^{1/2} \mu^{3/2}
+ {\hbar^2\over 12\pi m} \Lambda^3
\nonumber
\\
&+& {\mu\over 2\pi} \Lambda
+ O({1\over \Lambda}) \; .  
\label{rigolin}
\eeqa
The term proportional to $\Lambda^3$ is
independent of $\mu$ and can then be ignored. The term proportional
to $\Lambda$ which depends on $\mu$ may be absorbed by redefining
the bare chemical potential appearing in the original zero-temperature
mean-field grand-potential. By defining
\beq
\mu_r = \mu - \delta \mu
\eeq
with
\beq
\delta \mu = {g \Lambda \over 2\pi} \; .
\eeq
Eq ($\ref{rigolin}$) becomes:

\beq
{\Omega\over L} = -{\mu_r^2\over 2\, g}
- {2\over 3\pi} ({m\over \hbar^2})^{1/2} \mu_r^{3/2}
+ {1\over \beta L} \sum_{{\bf q}} \ln{\left(1 - e^{-\beta E_q} \right)} \; ,  
\label{diabic}
\eeq
which indeed coincides  Eq. (\ref{regdim-1d})
obtained with dimensional regularization if one identifies $\mu_r$ with
the measured chemical potential.

For the three-dimensional interacting Bose gas
we set $D=3$ in Eq. (\ref{succhia1}) and after integration we find  
\beq
{\Omega_g^{(0)}\over L^3} = {8\over 15\pi^2} ({m\over \hbar^2})^{3/2} \mu^{3/2}
\Big(1 + {1\over 2} \sqrt{1+Z} \Big(3 Z^2 + Z - 2 \Big)^{3/2} \Big) \; ,  
\eeq
where $Z=\hbar^2\Lambda^2/(4m\mu)$.
The zero-temperature total grand potential is then given by
\beqa
{\Omega_{0}\over L^3} + {\Omega_g^{(0)}\over L} &=& -{\mu^2\over 2\, g}
+ {8\over 15\pi^2} ({m\over \hbar^2})^{3/2} \mu^{5/2}
- {7\hbar^2\over 1280 \sqrt{2} \pi m} \Lambda^5
\nonumber
\\
&+& {\mu \over 12\pi^2} \Lambda^3 - {m \mu^2\over 4\pi^2 \hbar^2} \Lambda
+ O({1\over \Lambda}) \; ,  
\label{totLam}
\eeqa
in the large-$\Lambda$ limit. The term proportional to $\Lambda^5$ is
independent of $\mu$ and  can be ignored. Since now there are two
relevant (containing $\mu$) divergent terms, we need to renormalize both
parameters $\mu$ and $g$ to absorb them into the mean-field form.
As reported by Schakel (Schakel, 2008), this renormalization
of the bare physical parameters is achieved by setting
\beqa
\mu_r = \mu - {g \, \Lambda^3 \over 12 \pi^2} \; .
\\
g_r = g - {m\, g^2\, \Lambda \over 2\pi^2 \hbar^2}
\label{reng3D}
\eeqa
so that the renormalized grand potential becomes
\beq
{\Omega\over L^3} = -{\mu_r^2\over 2\, g_r}
- {8\over 15\pi^2} ({m\over \hbar^2})^{3/2} \mu_r^{5/2}
+ {1\over \beta L^3} \sum_{{\bf q}} 
\ln{\left(1 - e^{-\beta E_q} \right)} \; ,  
\eeq
which is equivalent to the equation of state Eq. (\ref{regdim-3d})
previously obtained through dimensional regularization.

It is worth noticing that the result (\ref{reng3D}) obtained by redefining 
the parameters of the theory $\mu$ and $g$ so that Eq. (\ref{totLam}) 
takes the mean-field form, coincides with the expression obtained from 
Eq. (\ref{again0}) with a high wavenumber cutoff $\Lambda$.

\subsubsection{Repulsive Bose gas in 2D - MCR}

The momentum-cutoff regularization of the two-dimensional repulsive Bose
gas requires
a very careful analysis. We first set $D=2$ into Eq. (\ref{succhia1})
and after integration we find  
\beq
{\Omega_g^{(0)}\over L^2} = {m\over 4\pi\hbar^2} \mu^2
\left(  {2Z^3 + 3 Z^2 + Z \over \sqrt{Z^2+Z}}
- \ln{(2 \sqrt{Z})}
\right) \; ,  
\label{caterina}
\eeq
where $Z=\hbar^2\Lambda^2/(4m\mu)$. The zero-temperature total grand
potential is then given by
\beqa
{\Omega_{0}\over L^2}+{\Omega_g^{(0)}\over L^2}
&=& - {\mu^2\over 2\, g} + {m\over 16\pi\hbar^2} \mu^2
+ {\hbar^2\over 32\pi\, m} \Lambda^4 + {\mu\over 8\pi} \Lambda^2
\nonumber
\\
&-& {m\over 8\pi\hbar^2} \mu^2
\ln{\left({\hbar^2\Lambda^2\over m\, \mu}\right)}
+ O({1\over \Lambda^4}) \; ,  
\eeqa
in the large-$\Lambda$ limit. The term proportional to $\Lambda^4$ is
independent of $\mu$ and can be ignored. The terms proportional to
$\Lambda^2$ and $\ln{(\Lambda^2)}$ depend on $\mu$ and therefore they must
be properly treated. First of all, we separate terms proportional to
integer powers of $\mu$ from a term containing a
logarithmic dependence by writing
\beq
\ln{\left({\hbar^2\Lambda^2\over m\, \mu}\right)} =
\ln{\left({\hbar^2\Lambda^2\over m\, \epsilon_c}\right)} +
\ln{\left({\epsilon_c\over \mu}\right)}
\eeq
where the energy $\epsilon_c$ is completely arbitrary. Now we can absorb
the divergent terms by setting  
\beqa
\mu_r &=& \mu - \frac{{g \, \Lambda^2 \over 8 \pi}}{1-\frac{m g}
{4 \pi \hbar^2} \ln {\frac{\varepsilon_c}{\mu}}} \; .
\\
g_r &=& g - {m\, g^2 \over 4\pi \hbar^2}
\ln{\left({\hbar^2\Lambda^2\over m \epsilon_c}\right)} \; .
\eeqa
to obtain the renormalized total grand potential
\beq
{\Omega\over L^2} = -{\mu_r^2\over 2\, g_r} - {m\over 8\pi\hbar^2} \mu_r^2
\ln{\left({\epsilon_c\over e\, \mu_r } \right)}\,  +
{1\over \beta L^2} \sum_{{\bf q}} \ln{\left(1 - e^{-\beta E_q} \right)} \; ,  
\eeq
which can finally be rewritten as
\beq
{\Omega\over L^2} = - {m\over 8\pi\hbar^2} \mu_r^2
\ln{\left({\epsilon_0\over \mu_r } \right)} +
{1\over \beta L^2} \sum_{{\bf q}} \ln{\left(1 - e^{-\beta E_q} \right)} \; ,
\label{vietto}
\eeq
where
\beq
\epsilon_0 = \epsilon_c \, e^{4\pi \hbar^2/(g_r m)-1/2} \; .
\label{vietto-bis}
\eeq
Again  the renormalized coupling constant  coincides with the expression 
obtained from  the scattering theory, Eq. (\ref{again0}), with a high 
wavenumber cutoff $\Lambda$.

Also in this two-dimensional case the total grand potential obtained with the
cutoff regularization, Eq. (\ref{vietto}), is the same as the one found with
dimensional-regularization, Eq. (\ref{cicredi}).

\subsection{Convergence-factor regularization (CFR)}

We now analyze a third method of regularization of zero-point fluctuations.
This method,  mainly used in condensed-matter physics
(Nagaosa, 1999; Stoof {\it et al.}, 2009; Altland and Simons, 2010),  
is based on the use of a convergence factor $e^{i \omega_n 0^+}$
when performing the Matsubara sums in Eqs. (\ref{pl0}) and (\ref{pl}).
As we shall see, in the case of the ideal Bose gas the presence of the
convergence factor removes completely the zero-point energy of quantum 
fluctuations, while in the case of the interacting Bose gas  
the cancellation of the zero-point divergence is
only partial.

\subsubsection{Ideal Bose gas - CFR}

As explained in detail by Altland and Simons (2010), the inclusion
of a convergence factor $e^{i \omega_n 0^+}$ into Eq. (\ref{pl0})  
exactly produces, after the complex integration associated to
the Matsubara sum, a counterterm which removes the zero-point divergence,
namely Eq. (\ref{flavione0}) is modified into 
\beq
{1\over 2\beta}
\sum_{n=-\infty}^{+\infty} \ln{[\beta^2(\hbar^2\omega_n^2+\xi_{q}^2)]} \,
e^{i\omega_n 0^+} = {1\over \beta } \ln{(1-e^{-\beta\, \xi_q})} \; ,  
\label{caa}
\eeq
where $e^{i \omega_n 0^+}$ means $\lim_{\delta\to 0^+}e^{i \omega_n \delta}$.
Consequently, the grand potential is given by Eq. (\ref{submarcos}),
as it must.

\subsubsection{Repulsive Bose gas in 1D and 3D - CFR}

In the case of the repulsive Bose gas the introduction of a convergence-factor
into Eq. (\ref{pl}) leads to \citep{diener2008}
\beqa
{1\over 2\beta}
\sum_{n=-\infty}^{+\infty} \ln{[\beta^2(\hbar^2\omega_n^2+E_{q}^2)]} \,
e^{i\omega_n 0^+}
\nonumber
\\
= {1\over 2}\left( {E_q\over 2} - {\hbar^2q^2\over 2m} - \mu \right)
+ {1\over \beta } \ln{(1-e^{-\beta\, \xi_q})} \; .  
\label{caa1}
\eeqa
As a consequence, in the continuum limit the zero-temperature grand
potential reads
\beq
{\Omega_0\over L^D} +{\Omega_g^{(0)}\over L^D} = - {\mu^2\over 2\, g} +
{1\over 2} \int {d^D{\bf q}\over (2\pi)^D}
\left( E_q - {\hbar^2q^2\over 2m} - \mu \right) \; ,  
\label{cca2}
\eeq
where $E_q$ is given by Eq. (\ref{spettrob}).

In 1D, after integration one finds
\beq
{\Omega_g^{(0)}\over L} = -
{2\over 3\pi} ({m\over \hbar^2})^{1/2} \mu^{3/2} \, ,
\eeq
confirming the equation of state (\ref{regdim-1d})
obtained above by both the dimensional
and  momentum-cutoff regularization.

In the three-dimensional case the integral of Eq. (\ref{cca2}) is
ultraviolet divergent. As we have previously seen, this kind of
divergence can be fully removed by dimensional
regularization or, equivalently, by momentum-cutoff regularization. Moreover,
it can also be removed by taking into account scattering
theory at the second order, namely Eq. (\ref{again}), which in the
continuum limit and  $D=3$ may be written as
\beq
{1\over g_r}
= {1\over g} + \int {d^3{\bf q}\over (2\pi)^3}
{m\over \hbar^2 q^2} \;  
\label{again1}
\eeq
with $g_r=4\pi\hbar^2 a_B/m$.
The integral in the right hand side of Eq. (\ref{again1}) is ultraviolet
divergent, but by inserting Eq. (\ref{again1}) into Eq. (\ref{cca2}) 
one obtains
\beqa
{\Omega_0\over L^3} &+&{\Omega_g^{(0)}\over L^3} = - {\mu^2\over 2\, g_r}
\nonumber
\\
&+& {1\over 2} \int {d^3{\bf q}\over (2\pi)^D}
\left( E_q - {\hbar^2q^2\over 2m} - \mu + {m\, \mu^2 \over \hbar^2 q^2}
\right) \; .  
\eeqa
This integral is now finite and the zero-temperature grand potential
once again is found to be
\beq
{\Omega_0\over L^3} + {\Omega_g^{(0)}\over L^3} = - {\mu^2\over 2\, g_r}
+ {8\over 15 \pi^2} ({m\over \hbar^2})^{3/2} \mu^{5/2} \; .
\eeq

\subsubsection{Repulsive Bose gas in 2D - CFR}

Also in the two-dimensional case the integral of Eq. (\ref{cca2}) is
ultraviolet divergent. Proceeding as in the 3D case with scattering theory
at the second order, i.e. by using the
2D version of Eq. (\ref{again1}) given by
\beq
{1\over g_r}= {1\over g} + \int {d^2{\bf q}\over (2\pi)^2}
{m\over \hbar^2 q^2} \; ,  
\label{again2}
\eeq
leads to the zero-temperature grand potential
\beqa
{\Omega_0\over L^2} &+&{\Omega_g^{(0)}\over L^2} = - {\mu^2\over 2\, g_r}
\nonumber
\\
&+& {1\over 2} \int {d^2{\bf q}\over (2\pi)^D}
\left( E_q - {\hbar^2q^2\over 2m} - \mu + {m\, \mu^2 \over \hbar^2 q^2}
\right) \; .  
\eeqa
Here the ultraviolet divergence has been cancelled  but the last term
introduces an infrared divergence which must be cured by introducing
a low-energy cutoff $\epsilon_c$ (and its corresponding wavenumber
$k_c=(m\epsilon_c)^{1/2}/\hbar$). In this way,  we find
\beq
{\Omega_g^{(0)} \over L^2} = {m\over 4\pi \hbar^2} \mu^2
\left( {1\over 4} - \ln{2} - {1\over 2} \ln{\frac{\epsilon_c}{4\mu}}
\right) \; ,
\eeq
which, after  some algebraic manipulations,
gives back the total grand potential
of Eq. (\ref{vietto}), with Eq. (\ref{vietto-bis}) and $\mu$
instead of $\mu_r$.

In conclusion, also for the two-dimensional interacting
Bose gas the three methods of regularization
(dimensional, momentum-cutoff, and convergence-factor)
of divergent Gaussian fluctuations
give rise to the same equation of state.
However, in 2D the convergence-factor regularization plus  scattering theory
needs also a further momentum-cutoff regularization of the residual
infrared divergence. 

\section{Experiments vs theory for bosonic superfluids}

\subsection{Repulsive Bose gas in 3D}

Despite the very large number of experiments with bosonic 
gases made of dilute and ultracold alkali-metal atoms, 
only in recent years zero-point energy (i.e. beyond-mean-field) 
effects have been measured. 

In 2008 \cite{papp2008} 
got informations on the zero-temperature chemical 
potential $\mu$ of a gas of $^{85}$Rb atoms as a function of 
the gas parameter $na_B^3$. Through the Feschbach resonance technique, they 
tuned the scattering length $a_B$ of their harmonically trapped 
atomic sample up to $1000\ a_0$ 
($a_0=0.53\cdot 10^{-8}$ cm is the Bohr radius). 
Since their sample had a mean number density $n=7.6\cdot 10^{-13}$ cm$^{-3}$ 
they were then varying the gas parameter $n {a_B}^3$ up to 0.002. 
Using two-photon Bragg spectroscopy to probe 
the Bogoliubov excitation spectrum, see Eq. (\ref{spettrob}), 
of the strongly interacting 3D Bose-Einstein condensate,
they then measured the energy required to promote an atom out of the 
condensate less the bare kinetic energy, namely:
\beq 
E_q - {\hbar^2q^2\over 2m} = 
\sqrt{{\hbar^2q^2\over 2m}\left({\hbar^2q^2\over 2m} + 2 \mu \right)} - 
{\hbar^2q^2\over 2m} \; . 
\eeq
In the regime of high-momentum excitations this quantity becomes 
\beq 
E_q - {\hbar^2q^2\over 2m} \simeq \mu  \; . 
\eeq
Thus, measuring $E_q - {\hbar^2q^2/(2m)}$ at high momenta \cite{papp2008} 
determined  the behavior of $\mu$ hence confirming the contributions of
quantum fluctuations as discussed above.

In fact, by using the one-loop grand potential 
given by (\ref{regdim-3d}) and the thermodynamic formula 
(\ref{number-find-fla}) one immediately finds at zero temperature 
\beq 
n = {\mu \over g} - {4\over 3\pi^2} 
\left({m\over \hbar^2}\right)^{3/2} {\mu^{3/2}} \; . 
\label{tegico}
\eeq
>From this equation one can easily  
determine $\mu$ as a function of $n$ and $a_B$, that is 
\beq
\mu = \mu_0 + \mu_{g}
\eeq 
where 
\beq 
\mu_0 = {4\pi\hbar^2\over m} a_B \, n 
\eeq
is the mean-field result with $g=4\pi\hbar^2a_B/m$ and 
\beq 
\mu_{g} = {4\pi\hbar^2\over m} a_B \, n \, \big( 
{\frac{32}{\sqrt \pi}} (n a_B^3)^{1/2} \big)  \;  
\label{mugiog}
\eeq
is the zero-temperature Gaussian correction, 
under the condition $n a_B^3 \ll 1$. 

\begin{figure}[t]
\centerline{\epsfig{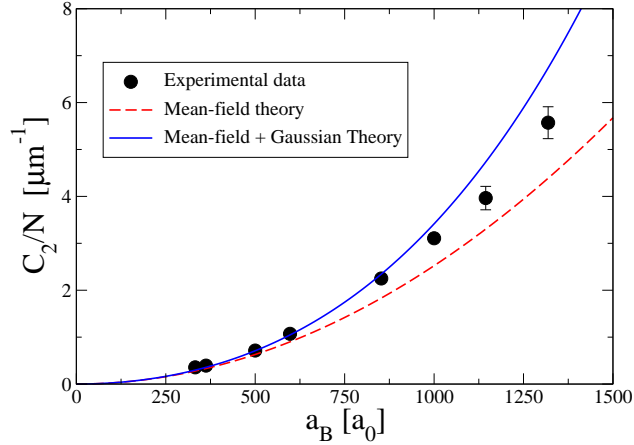}}
\small
\caption{Contact $C_2$ vs scattering length $a_B$ for the  
zero-temperature repulsive Bose gas. Comparison 
between mean-field theory (dashed line) 
and one-loop theory, i.e mean-field plus Gaussian 
fluctuations (solid line). Filled circles are experimental 
data of \cite{wild2012}, obtained with a $^{85}$Rb atomic 
condensate using rf spectroscopy, where 
$n=5.8\cdot 10^{12}$ cm$^{-3}$ is the mean number density 
and the gas parameter $na_B^3$ reaches a maximum of $0.0018$. 
The scattering length $a_B$ is in units of the Bohr 
radius, $a_0=0.53\cdot 10^{-8}$ cm. }
\label{fig3}
\end{figure}

In 2012 \cite{wild2012} found beyond-mean-field effects in the 
zero-temperature equation of state of a repulsive 3D Bose gas 
made of $N\simeq 6\cdot 10^4$ $^{85}$Rb atoms by measuring the 
two-body contact $C_2$. $C_2$ is an extensive thermodynamic quantity 
that is connected to the derivative of the total energy $E$ 
of the system with respect to the s-wave scattering length $a_B$ 
(\cite{tan2008}), namely 
\beq 
C_2= {8\pi m a_B^2\over \hbar^2} {dE\over da_B} \; . 
\label{gborghese0}
\eeq
The contact $C_2$ characterizes the tail of the 
momentum distribution of a many-body system of identical particles 
(\cite{tan2008}) and also the tail of the rate $\Gamma(\omega)$ 
for inducing transitions between spin states 
in rf spectroscopy (\cite{wild2012}). 
Indeed, in the experiment a pulse of radio-frequency $\omega$ 
was used to drive a Zeeman transition and to transfer 
a small fraction of spin-polarized bosonic atoms into 
another spin state. Then, from the observed $\Gamma(\omega)$,  
\cite{wild2012} extracted the value of the two-body contact $C_2$. In Fig. 
\ref{fig3} we plot their experimental data for $C_2 \over N$ vs $a_B$ 
as filled circles. In the figure the dashed line is the 
mean-field value of $C_2 \over N$, that is 
\beq 
{C_2\over N} = 16 \pi^2 n a_B^2 \; , 
\label{gborghese1}
\eeq
while the solid line is the one-loop value of $C_2 \over N$, given by 
\beq 
{C_2\over N} = 16 \pi^2 n a_B^2 \left( 1 + {5\over 2} {128\over 15\sqrt{\pi}} 
\sqrt{n a_B^3} \right) \; . 
\label{gborghese2}
\eeq
Both Eq. (\ref{gborghese1}) and  Eq. (\ref{gborghese2}) are 
obtained from Eq. (\ref{gborghese0}) and $E=\Omega + \mu N$ 
with the pertinent $\Omega$. As noticed in \cite{wild2012}, 
Fig. \ref{fig3}, confirms that by increasing the scattering 
length $a_B$ the two-body contact $C_2$ deviates from the mean-field 
prediction and the one-loop theory (mean-field 
plus Gaussian fluctuations) better reproduces the experimental data.

\subsection{Repulsive Bose gas in 2D}

In 2011 the thermodynamics of a 2D Bose gas was investigated 
by \cite{dalibard2011} with $^{87}$Rb atoms. 
As in previous thermodynamic investigations of a 3D Bose gas 
of $^7$Li atoms (\cite{salomon2010}), the analysis of 
\cite{dalibard2011} was performed by analyzing {\it in situ} measurements 
of the density profiles $n({\bf r})$ to extract the pressure $P(\mu,T)$ 
of the equivalent homogeneous gas. 

The idea is the following: starting from 
Eq. (\ref{cicredi}), which is the 2D grand potential $\Omega$ 
of the uniform system, the pressure $P=-\Omega/L^2$ 
reads 
\beq 
P(\mu,T) = {m\over 8\pi\hbar^2} \ln{\left({\epsilon_0\over \mu} \right)} 
- {k_BT} \int {d^2{\bf q}\over (2\pi)^2} 
\ln{\left( 1 - e^{ -\sqrt{{\hbar^2q^2\over 2m}
({\hbar^2q^2\over 2m}+2\mu)}/(k_BT)}\right)} 
\label{dali-dadi}
\eeq
while the corresponding number density $n$ is given by 
\beq 
n(\mu,T) = \left({\partial P \over \partial \mu}\right)_{T} \; . 
\eeq
Using the local density approximation
\beq 
\mu \to  \bar{\mu} - U({\bf r}) \; , 
\eeq
where $U({\bf r})$ is the space-dependent confining potential 
and $\bar{\mu}$ is the chemical potential of the non homogeneous system, 
one gets the local number density $n({\bf r})$ as 
\beq 
n({\bf r}) = n(\mu=\bar{\mu}-U({\bf r}),T) \; . 
\eeq
Reversing this procedure, from the local number density $n({\bf r})$ 
of the non homogeneous system one gets the pressure 
of the homogeneous system. The experimental data of $P(\mu,T)$ 
extracted by \cite{dalibard2011} for the trapped gas of $^{87}$Rb atoms 
suggest that the finite-temperature contribution to the pressure 
given by Eq. (\ref{dali-dadi}) is fully reliable. However, 
due to difficulties to reach extremely low temperatures, in this 
experiment they were not able to test the zero-temperature 
logarithmic behavior of the 2D equation of state. This important 
experimental investigation has been recently performed by \cite{makhalov2014} 
on a 2D system of composite bosons made of bound pairs of $^6$Li atoms. 
We shall discuss in detail this experiment in Section 6.2, comparing 
it with our theoretical predictions for attractive fermions 
in the deep BEC regime of the 2D BCS-BEC crossover.

\subsection{Repulsive Bose gas in 1D} 

In 2004 Kinoshita et al. (\cite{kinoshita2004}) 
reported the observation of a one-dimensional gas of cold rubidium-87 atoms 
confined in a cigar shaped regions by two orthogonal strong light traps 
and moving almost freely in the third direction. In the experiment  
the 1D interaction strength $g_{1D}$ is modified by changing the width 
$a_{\bot}$ of the transverse harmonic confinement, 
as $g_{1D}=g_{3D}/(2\pi a_{\bot}^2)$ 
where, as said above, $g_{3D}=4\pi\hbar^2a_{3D}/m$. 
By changing the trap intensities and hence the atomic interaction 
strength $\gamma= (m/\hbar^2)(g_{1D}/n_{1D})$ 
the atoms were made to act either like a Bose-Einstein condensate, 
when $\gamma \ll 1$ or like a Tonks-Girardeau gas of impenetrable 
bosons when $\gamma \gg 1$.
Kinoshita et al. (\cite{kinoshita2004}) thus measured the total 1D energy
and the length of the gas. With no free parameters and over a wide range
of coupling strengths, their data fit the exact solution for the ground state
of a 1D Bose gas as found by \cite{lieb1963}. 

This experiment confirms that in the weak-coupling 
regime ($\gamma \ll 1$) the 1D system is a quasi-condensate 
very well described by the one-loop grand potential (\ref{regdim-1d}).

In the strong coupling regime, also studied with a completely different 
experimental setup by \cite{paredes2004}, the 1D system is a Tonks-Girardeau 
gas, whose grand potential is given by 
\beq   
{\Omega \over L} = - {2\sqrt{2} \over 3\pi} ({m\over \hbar^2})^{1/2} \mu^{3/2} 
\label{santamad1}
\eeq
and the corresponding chemical potential reads
\beq 
\mu = {\pi^2 \hbar^2\over 2 m} n^2 \; . 
\label{santamad2}
\eeq
The failure in this case of the one-loop Gaussian approximation, which 
at $T=0$ gives the grand potential $\Omega$ of Eq.(\ref{regdim-1d}) 
instead of Eq.(\ref{santamad1}), is not surprising since Gaussian 
results are expected to be fully reliable only in the week-coupling regime. 

\section{Functional integration for fermionic superfluids}

We have seen that after regularization,
the zero-point energy of the Gaussian quantum
fluctuations contributes  a non-trivial term to the equation
of state of  an interacting $D$-dimensional Bose superfluid.
Three  different regularization approaches (dimensional regularization,
momentum-cutoff regularization and
convergence-factor regularization) produce the same
finite result, which is however dependent on the
dimensionality of the system. Moreover,
at variance with both dimensional regularization
and momentum-cutoff regularization which are self-contained,
convergence-factor regularization explicitly needs
scattering theory (or, again,
dimensional or cutoff regularization).

Extremely interesting is the study of
the divergent zero-point energy of a $D$-dimensional
two-spin-component Fermi superfluid in the BCS-BEC crossover. 
As will be discussed in detail below, 
the crossover from the weakly-paired
Bardeen-Cooper-Schrieffer (BCS) state to the Bose-Einstein
condensate (BEC) of molecular dimers
has been experimentally achieved using ultracold
fermionic alkali-metal atoms a few years ago in 3D
\citep{greiner2003,chin2004} and quite recently recently also in a
two-dimensional configuration \citep{makhalov2014}.
In the gas of paired fermions there are two kinds of elementary excitations:
fermionic single-particle excitations with energy
\beq
E_{sp}(k)=\sqrt{\left({\hbar^2k^2\over 2m}-\mu\right)^2+\Delta_0^2} \; ,  
\label{ex-fermionic}
\eeq
where $\Delta_0$ is the pairing gap, and  bosonic collective
excitations with energy 
\beq
E_{col}(q) = \sqrt{{\hbar^2q^2\over 2m} \left( \lambda \ {\hbar^2q^2\over 2m}
+ 2 \ m \ c_B^2 \right)} \; ,
\label{ex-bosonic}
\eeq
where $\lambda \ne 0$ gives the first correction
to the familiar low-momentum dispersion
$E_{col}(q) \simeq c_B \hbar q$. Eq. (\ref{ex-bosonic}) is obtained 
in the limit of a small-wavenumber $q$ from Gaussian 
fluctuations \citep{randeria1990,marini1998}. 
Notice that for a given scattering length both $\lambda$ and $c_B$ depend on
the chemical potential $\mu$ and so does
the energy gap $\Delta_0$. As we shall see, after regularization,
the zero-point energy of these elementary excitations gives a relevant
contribution to the equation of state of the fermionic superfluid.

Starting from the familiar BCS Lagrangian density
of paired (attractive) fermions \citep{nagaosa1999}
\beq
\mathscr{L} = \bar{\psi}_{s} \left[ \hbar \partial_{\tau}
- \frac{\hbar^2}{2m}\nabla^2 - \mu \right] \psi_{s}
+ g \, \bar{\psi}_{\uparrow} \, \bar{\psi}_{\downarrow}
\, \psi_{\downarrow} \, \psi_{\uparrow}
\label{bcs-lagrangian}
\eeq
where  $\psi_s({\bf r},\tau)$ and $\bar{\psi}_s({\bf r},\tau)$ 
are Grassman variables describing the
fermionic field and $g<0$ is the strength of the s-wave inter-atomic
coupling, the partition function ${\cal Z}$ of the uniform 
fermionic system in a $D$-dimensional volume $L^D$, 
and with chemical potential $\mu$ reads
\beq 
{\cal Z} = \int {\cal D}[\psi_{s},\bar{\psi}_{s}] 
\ \exp{\left\{ -{1\over \hbar} \ S[\psi_{s},\bar{\psi}_{s}]  \right\} } \; , 
\eeq
where 
\beq 
S[\psi_{s},\bar{\psi}_{s}] = \int_0^{\hbar\beta} 
d\tau \int_{L^D} d^D{\bf r} \ \mathscr{L}(\psi_{s},\bar{\psi}_{s}) 
\eeq
is the Euclidean action functional. 
Through the exact Hubbard-Stratonovich transformation (\cite{nagaosa1999}) 
the Lagrangian density ${\cal L}$, quartic in the fermionic fields, 
can be rewritten as a quadratic form by introducing the 
auxiliary complex scalar field $\Delta({\bf r},\tau)$, namely 
\beq
\mathscr{L}_e =
\bar{\psi}_{s} \left[  \hbar \partial_{\tau}
- {\hbar^2\over 2m}\nabla^2 - \mu \right] \psi_{s}
+ \bar{\Delta} \, \psi_{\downarrow} \, \psi_{\uparrow}
+ \Delta \bar{\psi}_{\uparrow} \, \bar{\psi}_{\downarrow}
- {|\Delta|^2\over g} \; .  
\eeq
In this way the partition function ${\cal Z}$
of the fermionic system can be rewritten exactly as 
\beq 
{\cal Z} = \int {\cal D}[\psi_{s},\bar{\psi}_{s}]\, 
{\cal D}[\Delta,\bar{\Delta}] \ 
\exp{\left\{ - {S_e[\psi_s, \bar{\psi_s},
\Delta,\bar{\Delta}] \over \hbar} \right\}} \; , 
\eeq
where 
\beq 
S_e[\psi_s, \bar{\psi_s},\Delta,\bar{\Delta}] = \int_0^{\hbar\beta} 
d\tau \int_{{L^D}} d^D{\bf r} \ 
\mathscr{L}_e(\psi_s, \bar{\psi_s},\Delta,\bar{\Delta})
\eeq
is the the (exact) effective Euclidean action. Notice that 
now there is a functional integration also over $\Delta({\bf r},\tau)$. 
This is the price to pay for having an effective Lagrangian 
that is quadratic, instead of quartic, in the fermionic fields 
$\psi_s({\bf r},\tau)$ and $\bar{\psi_s}({\bf r},\tau)$. 

The effect of fluctuations of the field $\Delta({\bf r},t)$ around its
mean-field value $\Delta_0$ (the pairing gap) may be analyzed at the
Gaussian level by taking
\beq
\Delta({\bf r},\tau) = \Delta_0 +\eta({\bf r},\tau)  \; ,
\eeq
where $\eta({\bf r},\tau)$ is the complex pairing field of bosonic
fluctuations. In particular, we are interested in the one-loop 
grand potential $\Omega$, given by 
\beq 
\Omega = - {1\over \beta} \ln{\left( {\cal Z} \right)} \simeq - {1\over \beta} 
\ln{\left( {\cal Z}_{mf} {\cal Z}_g \right)} = \Omega_{mf} + \Omega_{g} 
\; , 
\eeq
where 
\beq 
{\cal Z}_{mf} = \int {\cal D}[\psi_{s},\bar{\psi}_{s}]\, 
\exp{\left\{ - {S_e[\psi_s, \bar{\psi_s}, 
\Delta_0] \over \hbar} \right\}} \;  
\eeq
is the mean-field partition function and 
\beq
{\cal Z}_g = \int {\cal D}[\psi_{s},\bar{\psi}_{s}]\, 
{\cal D}[\eta,\bar{\eta}] \ 
\exp{\left\{ - {S_g[\psi_s, \bar{\psi_s},
\eta,\bar{\eta},\Delta_0] \over \hbar} \right\}} 
\eeq
is the partition function of Gaussian pairing fluctuations, i.e. 
neglecting cubic and quartic contributions of $\eta$. Thus, 
one may write the total one-loop grand potential as
\beq
\Omega = \Omega_{mf} + \Omega_{g}.
\label{totom}
\eeq
In Eq. (\ref{totom})
\beq
\Omega_{mf} = \Omega_0 + \Omega_{mf}^{(0)} + \Omega_{mf}^{(T)}
\eeq
is the so-called mean-field grand potential, which includes
the grand potential of the order parameter $\Delta_0$
\beq
\Omega_{0} = - {\Delta_0^2\over g}\, L^D\; ,
\label{mf-delta0}
\eeq
the zero-point energy of fermionic single-particle excitations 
\beq
\Omega_{mf}^{(0)} = - \sum_{\bf k} E_{sp}(k) \; ,
\label{mf-fermionic}
\eeq
and the finite-temperature grand potential
of the fermionic single-particle excitations
\beq
\Omega_{mf}^{(T)} = {2\over \beta }
\sum_{\bf k} \ln{(1+e^{-\beta\, E_{sp}(k)})} \; .  
\eeq
In addition,
\beq
\Omega_{g} = \Omega_{g}^{(0)} + \Omega_{g}^{(T)} \; ,  
\eeq
is the grand potential of the bosonic Gaussian fluctuations, which 
includes the zero-point energy of bosonic collective excitations 
\beq
\Omega_{g}^{(0)} = {1\over 2} \sum_{\bf q} E_{col}(q) \; ,
\label{omegacol-div-miomio}
\eeq
and their finite-temperature contribution
\beq
\Omega_{g}^{(T)} = {1\over \beta }
\sum_{\bf q} \ln{(1- e^{-\beta\, E_{col}(q)})} \; .  
\label{malu}
\eeq

Clearly both $\Omega_{mf}^{(0)}$ and
$\Omega_{g}^{(0)}$ are ultraviolet divergent in the dimensions $D=1,2,3$.
Regularization of these divergent terms
is now complicated by the presence of
the BCS-BEC crossover. Very recently we have obtained interesting analytical
results in the BEC regime of the BCS-BEC crossover by removing
the divergences of single-particle and collective excitations
both in 3D \citep{sala-giacomo} and 2D
\citep{sala-flavio}. We shall discuss
the key ideas of these calculations in the next two subsections.

\subsection{Three-dimensional attractive Fermi gas}

Scattering theory plays an essential role
in the description of a three-dimensional attractive Fermi gas
which undergoes BCS-BEC crossover.
As recalled above, the bare interaction
strength $g$ appearing in the  Lagrangian density (\ref{bcs-lagrangian})
is related to the physical s-wave scattering length $a_F$ of fermions by
\beq
{m\over 4\pi\hbar^2 a_F} = {1\over g} +
{1\over {L^3}} \sum_{|{\bf k}|<\Lambda} {m\over \hbar^2k^2 } \; ,  
\label{magic}
\eeq
where, as usual, the ultraviolet cutoff $\Lambda$ is introduced to avoid the
divergence of the second term on the right side. 
We recall at this point that the low energy scattering length is
negative for an attractive potential if no bound states are present,
while it becomes positive when the interaction is so attractive as
to admit a bound state.
Eq. (\ref{magic}) allows for the change of sign of $a_F$ as the strength
of the attractive potential becomes more and more negative.
In fact, in the continuum limit
$\sum_{\bf k}\to L^3\int d^3{\bf k}/(2\pi)^3$,
after integration over momenta, it reads
\beq
{m\over 4\pi\hbar^2 a_F} = {1\over g} +  
{m\over 2\pi^2\hbar^2} \Lambda \; .
\label{magic2}
\eeq
Therefore,
in the weak-coupling BCS limit, where $g\to 0^-$, the first term
on the r.h.s. of Eq. (\ref{magic2}) dominates
and $a_F=mg/(4\pi\hbar^2)\to 0^-$, while 
in the strong-coupling BEC limit, where $g\to -\infty$, the second
term on the r.h.s. of Eq. (\ref{magic2}) dominates and $a_F=\pi/(2\Lambda)
\to 0^+$ when $\Lambda$ is sent to infinity \citep{gurarie2007,schakel2008}. 

\begin{figure}[t]
\centerline{\epsfig{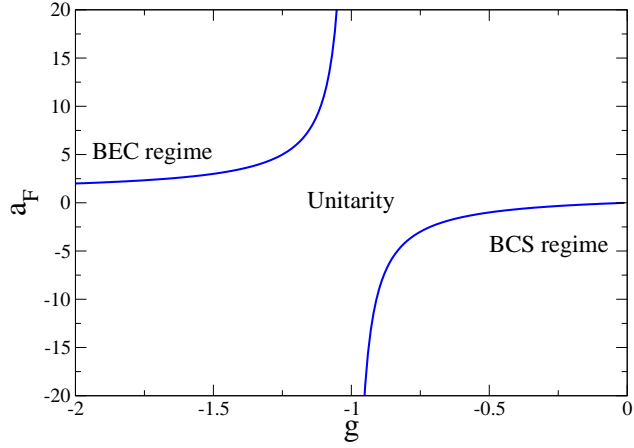}}
\small
\caption{Fermionic scattering length $a_F$ as a function 
of the bare interaction strength $g$, for a finite ultraviolet
cutoff $\Lambda$, see Eq. (\ref{magic2}). 
In the plot $a_F$ is in units of $m/(4\pi \hbar^2)$ 
and $m\Lambda/(2\pi^2\hbar^2) = 1$.}
\label{fig4}
\end{figure}

In addition, from Eq. (\ref{magic2}) one finds that 
a bound state, that is a pole in the T-matrix of Eq. (\ref{again0}),  
is possible only if $g< -2\pi^2\hbar^2/(m\Lambda)$. 
See Fig. \ref{fig4}. In the presence of a bound state of energy $-\epsilon_B$, 
for the T-matrix one has 
\beq 
{1\over T(-\epsilon_B )} = 0 \; , 
\eeq
and consequently from Eq. (\ref{again}) it follows 
\beq 
- {1\over g} = {1\over {L^3}} \sum_{|{\bf k}|<\Lambda} {m\over \hbar^2k^2 + 
m \epsilon_B } \; . 
\eeq
After integration over momenta one obtains 
\beq 
- {1\over g} = {m\over 2\pi^2 \hbar^2} 
\left( \Lambda 
-\sqrt{{m \epsilon_B\over \hbar^2}} 
\arctan{\left( {\Lambda \over \sqrt{{m \epsilon_B\over \hbar^2}}} \right)} 
\right)  \; , 
\label{magic3}
\eeq
and comparing Eq. (\ref{magic2}) with Eq. (\ref{magic3}) one finds 
\beq 
a_F = \sqrt{{\hbar^2 \over m \epsilon_B}} {\pi \over 2 
\arctan{\left( {\Lambda \sqrt{{\hbar^2 \over m \epsilon_B}}} \right)} } \; . 
\eeq
In the limit $\Lambda \to +\infty$ from this interesting formula 
one gets 
\beq 
\epsilon_B = {\hbar^2\over m a_F^2} \; ,
\eeq
that is the familiar relation between the bound-state energy 
and the 3D s-wave scattering length $a_F$. 

In 2015 \cite{sala-giacomo} found the regularized 
zero-temperature grand potential of the fermionic superfluid 
in the deep BEC regime in the form 
\beq
\Omega =
- L^3 {(1+\alpha) \over 256 \pi}
\left({2m\over \hbar^2}\right)^{3/2} {\Delta_0^4\over |\mu|^{3/2}} \;  
\label{omega-bec-miomio}
\eeq
with $\alpha=2$ due to zero-point Gaussian fluctuations, 
performing cutoff regularization and renormalization of
Gaussian fluctuations (with a procedure very similar to
the one discussed in Section IV.B for three-dimensional
interacting bosons), and taking into account the result
\beq
\Lambda = {\pi \over 2\, a_F} \;  
\label{socherompera}
\eeq
from Eq. (\ref{magic2}) when $g\to -\infty$. 
Let us briefly discuss the derivation of Eq. (\ref{omega-bec-miomio}).
In the deep BEC regime of the crossover, where the fermionic scattering
length $a_F$ becomes positive, the chemical potential $\mu$ becomes
negative and the regularized zero-temperature
mean-field grand potential reads \citep{sala-giacomo}
\beq
\Omega_{mf} =
- L^3 {1\over 256 \pi}
\left({2m\over \hbar^2}\right)^{3/2} {\Delta_0^4\over |\mu|^{3/2}} \; .  
\label{omega-bec00}  
\eeq
This expression  may be obtained following Leggett seminal paper
on the BEC-BCS crossover \citep{leggett1980} by using
the convergence-factor regularization supplemented by Eq. (\ref{magic})
and letting $\Lambda \to \infty$.
It may also  be obtained easily through dimensional regularization or
momentum-cutoff regularization, using procedures which are very similar
to the ones discussed in previous sections for bosons.

Regularization of the contribution to $\Omega$ from the bosonic Gaussian
fluctuations $\Omega_{g}^{(0)}$ is more delicate. In fact, expanding
the cutoff-regularized zero-temperature Gaussian grand potential,
Eq. (\ref{omegacol-div-miomio}), in powers of the momentum-cutoff
$\Lambda$ one finds at zero temperature \citep{sala-giacomo}
\beqa
{\Omega_{g}^{(0)}\over L^3} &=&
{\hbar^2 \lambda^{1/2}\over 40\pi^2 m} \Lambda^5
+ {m c_B^2 \over 12\pi^2 \lambda^{1/2}}\Lambda^3  
\nonumber
\\
&-& {m^3c_B^4\over 4\pi^2 \lambda^{3/2}} \Lambda +
{8 m^4c_B^5\over 15\pi^2 \hbar^2 \lambda^{2}} + O({1\over \Lambda}) .
\label{piopiopio}
\eeqa
The term proportional to $\Lambda^5$ can be neglected since
it does not depend on $\mu$ in the deep BEC limit where $\lambda=1/4$. 
In the same limit the term proportional to
$\Lambda^3$ can be absorbed by renormalizing $\mu$. Notice that the term
proportional to $\Lambda$ is finite  because in the BEC limit
$c_B^4$ goes to zero faster than $1/\Lambda$ \citep{sala-giacomo}. 
The same result is obtained by using the convergence-factor regularization
which gives an additional negative energy
$-(\lambda^{1/2}\hbar^2q^2/(2m) + mc_B^2/\lambda^{1/2})$ 
in the bosonic Gaussian 
grand potential $\Omega_{g}^{(0)}$ of Eq. (\ref{omegacol-div-miomio})
that exactly cancels the terms
proportional to $\Lambda^5$ and $\Lambda^3$ of Eq. (\ref{piopiopio}).  

Remarkably, the term of Eq. (\ref{piopiopio})
which is independent of the momentum cutoff $\Lambda$, i.e
\beq
\Omega_{g}^{(\Lambda-independent)}
= L^3 \, {8 m^4c_B^5 \over 15\pi^2 \hbar^2 \lambda^{2} } \; ,
\label{gauss-solodim}
\eeq
in the deep BEC regime is subleading with respect to the cutoff-regularized
term proportional to $\Lambda$. 
Thus, in the deep BEC regime, where
$|\mu|=\hbar^2/(2ma_F^2)$, $\lambda=1/4$ and $mc_B^2=\Delta_0^2/(8|\mu|)$
(notice that $\lambda$ and $c_B$ appear in the collective bosonic
excitations of Eq. (\ref{ex-bosonic})), the leading convergent
term of Eq. (\ref{piopiopio}) is
\beq
\Omega_{g} = - L^3 \ {\alpha \over 256 \pi}
\left({2m\over \hbar^2}\right)^{3/2} {\Delta_0^4\over |\mu|^{3/2}} \; ,  
\label{omegacol-bec11}
\eeq
with $\alpha =2$ \citep{sala-giacomo}. 
Even more remarkable is the
fact that performing dimensional regularization of Eq.
(\ref{omegacol-div-miomio}) one gets directly Eq. (\ref{gauss-solodim}).
In other words, dimensional regularization
of Eq. (\ref{omegacol-div-miomio}) does not give the same result
of momentum-cutoff regularization and convergence-factor regularization
because in our momentum-cutoff (or convergence-factor) regularization
$\Lambda$ is  constrained by Eq. (\ref{socherompera}). Clearly,
within our momentum-cutoff (or convergence-factor)
regularization scheme, from Eqs.
(\ref{omega-bec00}) and (\ref{omegacol-bec11}) one immediately
obtains Eq. (\ref{omega-bec-miomio}). 

In conclusion, in the deep BEC regime the leading term of
the zero-temperature one-loop grand potential can be written as 
\beq 
\Omega = \Omega_{mf} + \Omega_{g} = - L^3 (1+\alpha) {m\over 2 \pi \hbar^2 a_F} 
(\mu + {1\over 2} \epsilon_B)^2 \; , 
\label{dpdpdp}
\eeq
having taken into account the result \citep{diener2008,sala-giacomo}
\beq
\mu = -{1\over 2} \epsilon_B + {1\over 4} {\Delta_0^2\over \epsilon_B} \; 
\eeq 
with $\epsilon_B=\hbar^2/(m a_F^2)$ derived in the BEC regime ($a_F\to 0^+$) 
from the gap equation
\beq
\left( {\partial \Omega_{mf}\over \partial \Delta_0}\right)_{L^3,\mu}
= 0 \; .
\eeq
Eq. (\ref{dpdpdp}) is the familiar grand potential 
\beq 
\Omega = L^3 {m_B \over 8 \pi \hbar^2 a_B} \mu_B^2 
\eeq
of weakly-interacting repulsive composite bosons 
of mass  $m_B=2m$, density $n_B=n/2$, chemical potential 
$\mu_B=2(\mu + \epsilon_B/2)$, and the boson-boson 
scattering length 
\beq
a_B= {2\over (1+\alpha)} = {2\over 3} \ a_F \; .  
\label{akbar}
\eeq
This result is in good agreement with other beyond-mean-field
theoretical predictions: $a_B/a_F \simeq 0.75$
based on a diagrammatic approach \citep{pieri2000},
$a_B/a_F \simeq 0.60$ derived from a four-body analysis \citep{petrov2004}
and also from Monte Carlo simulations \citep{astra2004},
and $a_B/a_F\simeq 0.55$ obtained with convergence
factors \citep{hu2006,diener2008}. However,
contrary to all other predictions wich are based at some points 
on numerical calculations, our result, based on a transparent 
cutoff regularization and subsequent renormalization of bare physical 
parameters \citep{sala-giacomo}, is fully analytical.

\subsection{Two-dimensional attractive Fermi gas}

In the analysis of the two-dimensional attractive Fermi gas
one must remember that, contrary to the three-dimensional case,
two-dimensional realistic interatomic potentials for alkali atoms 
always exhibit a bound state 
and correspondingly a positive two-dimensional s-wave scattering
length \citep{randeria1990,marini1998,bertaina2011}.
In particular, according to Mora and Castin \citep{mora2009}
the binding energy $\epsilon_b>0$ of two fermions can be written
in terms of the positive two-dimensional fermionic scattering length $a_F$ as
\beq
\epsilon_b= {4\over e^{2\gamma}}{\hbar^2\over m a_F^2} \; ,  
\label{eb-af}
\eeq
where $\gamma=0.577...$ is the Euler-Mascheroni constant. Moreover,
the attractive (negative) interaction strength $g$ of s-wave pairing
is related to the
binding energy $\epsilon_b>0$ of a fermion pair in vacuum by
the expression \citep{randeria1989}
\beq
- \frac{1}{g} = \frac{1}{2L^2} \sum_{\bf k} \frac{1}{{\hbar^2k^2\over 2m} +
\frac{1}{2} \epsilon_b} \; .
\label{g-eb}
\eeq
Randeria {\it et al.} \citep{randeria1990} showed that
in the two-dimensional BCS-BEC crossover,
at zero temperature ($T=0$) the mean-field grand
potential $\Omega_{mf}$ can be written as
\beq
\Omega_{mf} = - {m L^2\over 2\pi \hbar^2}
(\mu + {1\over 2} \epsilon_b )^2  \;  
\label{omega-mf-pio}
\eeq
taking into account that the zero-temperature
2D gap equation gives \citep{randeria1990}
\beq
\Delta_0=\sqrt{2\epsilon_b(\mu+\epsilon_b/2)} \; .  
\eeq
By using
\beq
n = - {1\over L^2} {\partial \Omega_{mf}\over \partial \mu}
\eeq
one immediately finds the chemical potential $\mu$ as a
function of the number density $n=N/L^2$, i.e.
\beq
\mu = {\pi \hbar^2 \over m} n - {1\over 2} \epsilon_b \; .
\label{echem-mf-pio}
\eeq
In the BCS regime, where $\epsilon_b \ll \epsilon_F$, with 
$\epsilon_F=\pi\hbar^2n/m$, 
one finds $\mu \simeq \epsilon_F >0$ while in the BEC regime,
where $\epsilon_b \gg \epsilon_F$ one has
$\mu \simeq - \epsilon_b/2 <0$. Clearly, both
Eqs. (\ref{omega-mf-pio}) and (\ref{echem-mf-pio})
do not reproduce the expected logarithmic behavior in the deep BEC regime,  
where there should be a two-dimensional Bose gas of repulsive
composite bosons \citep{schick1971,popov1972}.

By performing dimensional regularization of Gaussian fluctuations
(with a procedure that is very similar to the one discussed in Section III.C
for two-dimensional interacting bosons),
we have recently found \citep{sala-flavio}
that the zero-temperature grand potential becomes
\beq
\Omega = - {m L^2\over 64\pi\hbar^2}  
(\mu + {1\over 2}\epsilon_b)^2 \ \ln{\left({\epsilon_b\over
2 (\mu + {1\over 2}\epsilon_b) } \right)} \; ,  
\label{finiamolaqui}
\eeq
in the deep BEC regime of this two-dimensional Fermi superfluid,
where the chemical potential $\mu$ becomes
negative and $\lambda$ of Eq. (\ref{ex-bosonic}) goes to $1/4$.

Let us briefly discuss the derivation of Eq. (\ref{finiamolaqui}).
Setting $g_0=\pi\hbar^2/m$ and $\mu_0=\mu+\epsilon_b/2$, the 2D
zero-temperature mean-field grand potential (\ref{omega-mf-pio})
can be written as
\beq
\Omega_{mf} = - L^2 {\mu_0^2 \over 2\, g_0} \; .
\label{finiamola1}
\eeq
In addition, the Bogoliubov's sped of sound $c_B$ 
which appears in Eq. (\ref{ex-bosonic})
satisfies the relation \citep{marini1998,sala-flavio}
\beq
m c_B^2 = \mu + {1\over 2} \epsilon_b = \mu_0 \; ,
\label{stoigata}
\eeq
and consequently the 2D zero-temperature Gaussian grand potential
becomes
\beq
\Omega_g = L^2 \int {d^2{\bf q}\over (2\pi)^2}
\sqrt{{\hbar^2q^2\over 2m} \left( \lambda \ {\hbar^2q^2\over 2m}
+ 2 \, \mu_0 \right)} \;    
\label{finiamola2}
\eeq
with $\lambda =1/4$ in the deep BEC regime (Salasnich and Toigo, 2015).
Quite remarkably, the two-dimensional Eqs. (\ref{finiamola1})
and (\ref{finiamola2}) for the attractive Fermi gas
are formally equivalent to Eqs. (\ref{omega-mf}) and (\ref{zz})
of the $D$-dimensional repulsive Bose gas when $D=2$
(apart for the value of $\lambda$ that is equal to one for repulsive bosons).
Thus, one can use one of the three regularization procedures
discussed in the previous sections to get the zero-temperature
regularized total grand potential in the deep BEC regime
(where $\lambda=1/4$), that is Eq. (\ref{finiamolaqui}).
In our recent paper \citep{sala-flavio} we have used dimensional
regularization, which has the advantage of being independent
of scattering theory.

Introducing $\mu_B = 2(\mu + \epsilon_b/2)$ as the chemical potential
of composite bosons with mass $m_B=2m$ and
density $n_B=n/2$, the zero-temperature
total grand potential (\ref{finiamolaqui}) can be rewritten as
\beq
\Omega = - {m_B L^2\over 8\pi\hbar^2}
\mu_B^2 \ \ln{\left({\epsilon_b\over \mu_B } \right)} \; .
\label{moana}
\eeq
As usual, the total density of bosons $n_B=n/2$ is obtained
in terms of $\mu_B=2(\mu+\epsilon_b/2) $ from the zero-temperature
thermodynamic formula
\beq
n = - {1\over L^2} {\partial \Omega\over \partial \mu} \; ,
\eeq
which leads to:
\beq
n_B = {m_B\over 4\pi \hbar^2} \mu_B
\ln{\left({\epsilon_b \over \mu_B \ e^{1/2}} \right)} \; .
\label{law}
\eeq
Inserting Eq. (\ref{eb-af}), which gives the binding energy $\epsilon_b$
of two fermions in terms of their s-wave scattering length $a_F$,
into Eq. (\ref{law}) we exactly recover Popov's 2D equation
of state \citep{popov1972} of weakly-interacting bosons
with scattering length $a_B$, i.e.
\beq
n_B = {m_B\over 4\pi \hbar^2} \mu_B
\ln{\left({4\hbar^2\over m_B \mu_B a_B^2 e^{2\gamma+1}}\right)} \; ,
\label{iopopov}
\eeq
provided that we identify the effective bosonic scattering
length $a_B$ with \citep{sala-flavio}:
\beq
a_B={1\over 2^{1/2}e^{1/4}} \ a_F \: .
\label{figata}
\eeq
Remarkably, the value $a_B/a_F= 1/(2^{1/2}e^{1/4}) \simeq 0.551$
from this analytical formula is in full agreement with
$a_B/a_F=0.55(4)$ obtained by Monte Carlo calculations
\citep{bertaina2011,bertaina2013} and $a_B/a_F=0.56$
very recently derived by using Gaussian fluctuations
with convergence-factor regularization \citep{he2015}.

\subsection{One-dimensional attractive Fermi gas}

Even if we know that the Gaussian approximation is inadequate to treat 
strongly interacting Bose systems in 1D (Lieb and Liniger, 1963), 
nonetheless for the sake of completeness in this subsection we analyze the
one-dimensional ($D=1$) attractive Fermi gas at zero temperature
taking into account the mean-field contributions of Eqs. (\ref{mf-delta0})
and (\ref{mf-fermionic}) due to fermionic single-particle
excitations (\ref{ex-fermionic}) and the Gaussian quantum
fluctuations of Eq. (\ref{omegacol-div-miomio}), due to bosonic collective
excitations (\ref{ex-bosonic}). The 1D problem of fermions
with contact attractive interaction was exactly solved in 1967 by
Gaudin using the Bethe ansatz \citep{gaudin1967}. Similarly to the
two-dimensional case, also for the 1D attractive Fermi gas
for any strength $g<0$ it exists a bound state
of energy $\epsilon_b$. The chemical potential $\mu$ is positive
in the BCS regime of weak attraction while it becomes negative and approaches
$-\epsilon_b/2$ in the Tonks-like regime of strong attraction
\citep{gaudin1967,fuchs2004}. We use the words ``Tonks-like'' because,
as we shall see, this strongly-attractive regime of 1D fermions
is actually a Tonks-Girardeau regime \citep{girardeau1960},
where there is no quasi-BEC but instead there is
a gas of strongly-repulsive 1D bosons \citep{gaudin1967,fuchs2004}.

More than twenty years ago Casas {\it et al.} \citep{casas1991}
have studied the zero-temperature 1D mean-field theory. 
>From their results one immediately finds that
in the deep Tonks-like regime ($(\mu + \epsilon_b/2)/\epsilon_b/2 \ll 1$)  
Eqs. (\ref{mf-delta0}) and (\ref{mf-fermionic}) give
\beq
\Omega_{mf} = - {L\over 2\epsilon_b^{1/2}} ({m\over \hbar^2})^{1/2}
\left( \mu + {1\over 2} \epsilon_b \right)^2
\eeq
where $\epsilon_B=mg^2/(4\hbar^2)$ is the binding energy
of fermionic pairs \citep{casas1991}.
Setting $g_0=-g = 2 (\hbar^2\epsilon_b/m)^{1/2}$ and
$\mu_0=\mu+\epsilon_b/2$, the 1D
zero-temperature mean-field grand potential can be rewritten as
\beq
\Omega_{mf} = -L {\mu_0^2 \over 2\, g_0} \; .
\label{1d-finiamola1}
\eeq
In addition, also in 1D the Bogoliubov's speed of sound $c_B$ 
which appears in Eq. (\ref{ex-bosonic}) 
satisfies the formula (\ref{stoigata}) 
and consequently the 1D zero-temperature Gaussian grand potential
becomes
\beq
\Omega_g = L \int_{-\infty}^{+\infty} {dq \over (2\pi)}
\sqrt{{\hbar^2q^2\over 2m} \left( \lambda \ {\hbar^2q^2\over 2m}
+ 2 \, \mu_0 \right)} \;    
\label{1d-finiamola2}
\eeq
again with $\lambda =1/4$ in the deep Tonks-like regime. Clearly,
Eqs. (\ref{1d-finiamola1})
and (\ref{1d-finiamola2}) of the attractive 1D Fermi gas
are formally equivalent to Eqs. (\ref{omega-mf}) and (\ref{zz})
of the $D$-dimensional repulsive Bose gas when $D=1$
(apart for $\lambda$, that is equal to one for repulsive bosons).
Thus, we can again use one of the three regularization procedures
discussed in the previous sections to get the zero-temperature
regularized total grand potential $\Omega=\Omega_{mf}+\Omega_g$
in the deep Tonks-like regime (where $\lambda=1/4$). In this way we find
\beq
{\Omega \over L} =  - ({m\over \hbar^2})^{1/2} \left[
{1\over 2\epsilon_b^{1/2}}
( \mu + {1\over 2} \epsilon_b )^2
+ {8\over 3 \pi}  
( \mu + {1\over 2} \epsilon_b)^{3/2} \right] \: ,  
\eeq
showing that in the deep Tonks-like regime
the mean-field contribution to the zero-temperature
grand potential is subleading with respect to the Gaussian one.
Using the thermodynamic relation
\beq
n = - {1\over L} {\partial \Omega\over \partial \mu}
\eeq
for the 1D number density $n=N/L$ of fermions and the leading Gaussian
term for the grand potential we find
\beq
\mu = - {1\over 2} \epsilon_b + {\pi^2\hbar^2\over 16 m} n^2
\label{esbagliata}
\eeq
in the deep Tonks-like regime of strong interaction, which
corresponds to the very dilute limit.
The term $-\epsilon_b/2$ of Eq. (\ref{esbagliata}) is
exactly the first term of Gaudin theory \citep{gaudin1967}
in a low-density series expansion in powers of $n$ \citep{casas1991}. The term
$\pi^2\hbar^2n^2/(16 m)$ of Eq. (\ref{esbagliata}) is similar
but not equal to the second term of the exact Gaudin expansion, that instead
gives $\pi^2\hbar^2n^2/(32 m)$ \citep{casas1991}.
Thus, in the one dimensional case Gaussian fluctuations
improve the mean-field theory but do not produce the correct
equation of state in the Tonks-like regime. 
We remind that a Tonks gas can also be obtained starting 
from a 1D repulsive Bose gas by strongly increasing 
its positive interaction strength \citep{lieb1963}. 

\section{Experiments vs theory for fermionic superfluids} 

\subsection{Attractive Fermi gas in 3D}

Beyond-mean-field effects in the frequencies of collective excitations
of a 3D fermionic superfluid under external confinement  were first 
predicted by \cite{stringari2004} and experimentally detected in a 
dilute gas of $^6$Li atoms 
at very low temperatures by \cite{grimm2004} and by \cite{grimm2007}. 
In these experiments the atomic gas 
was confined by an external anisotropic harmonic potential
\beq
U({\bf r}) = {m\over 2} \left( \omega_{\bot}^2 (x^2+y^2) 
            + \omega_z^2 z^2 \right) \; , 
\eeq
where $\omega_{\bot}$  and $\omega_z$ are the cylindric radial and 
longitudinal frequencies, respectively. 
The collective dynamics of the system is described efficiently by the 
hydrodynamic equations of superfluids (for a review see \cite{giorgini2008}), 
modified by the inclusion of the external potential $U({\bf r})$, namely 
\begin{eqnarray}
  {\partial n\over \partial t} + 
  {\boldsymbol \nabla} \cdot \left( n \, {\bf v} \right) 
  = 0 \; , 
  \label{dyn1n}
 \\
  m {\partial {\bf v}\over \partial t} + {\boldsymbol \nabla} 
  \left[{1\over 2} m v^2 + \mu[n,a_F] + U({\bf r})\right] = {\bf 0} \; .   
  \label{dyn2n}
\end{eqnarray}
where the zero-temperature equation of state is encoded in the 
explicit expression of the chemical potential $\mu$ as a function 
of the local density $n$ and of the s-wave scattering length $a_F$ of fermions.
For a 3D system of attractive fermions one may find analytical solutions 
of Eqs. (\ref{dyn1n}) and (\ref{dyn2n}) corresponding to the breathing 
collective modes,  both in the BEC ($a_F \rightarrow 0^+$) and BCS 
($a_F \rightarrow 0^-$) regimes and at unitarity ($|a_F| \rightarrow \infty$) 
(\cite{stringari2004}). As a matter of fact analytic expressions for the 
collective frequencies (see for instance \cite{giorgini2008}) may also 
be calculated if the equation of state is of the type $\mu = \mu_0\, 
n^{\gamma}$ for (polytropic equation of state).
For very elongated cigar--shaped traps 
($\omega_{\bot}/\omega_z \gg 1$) the collective radial 
breathing mode frequency $\Omega_{\bot}$ is given by 
\begin{equation}
\Omega_{\bot} = \sqrt{2(\gamma +1)} \, \omega_{\bot} \; , 
\end{equation}
while the collective longitudinal breathing mode $\Omega_z$ is 
\begin{equation} 
\Omega_{z} = \sqrt{3\gamma +2\over \gamma +1} \, \omega_{z} \; .  
\end{equation} 
\cite{nick2005} interpolated between the BEC (with Lee,Huang,Yang 
(\cite{lhy1957}) correction and BCS (with mean field interaction) through the 
unitary regime, by introducing an effective polytropic index $\gamma$ as the 
logarithmic derivative of the chemical potential $\mu$, that is
\begin{equation}
\gamma = {n\over \mu} {\partial \mu\over \partial n} \; .   
\label{poly}
\end{equation}
This approach (\cite{nick2005}) predicted relevant deviations from the 
mean-field results for the frequencies of collective breathing modes 
of a two-component Fermi gas of $^6$Li atoms  to unitarity ($a_F=\infty$), 
which were confirmed by the experiment of \cite{grimm2007}. 

A direct measurement of the equation of state of an attractive 
ultracold fermions system  was performed by \cite{navon2010} by 
absorption imaging an harmonically trapped sample of  $^6$Li atoms at 
ultralow temperatures. Parametrizing their data for the pressure 
vs. density with analytical functions \cite{navon2010} were able to 
extract relevant physical quantities, such as beyond mean-field 
corrections, for the superfluid system in the entire BCS-BEC crossover. 
In particular, in the BEC regime, the data are well reproduced 
by the equation of state of superfluid dilute composite bosons 
(\cite{leyronas2007}) confirming the coefficient of the \cite{lhy1957} 
term and allowed the first experimental determination of the 
scattering length between composite bosons in terms of scattering 
length between fermions as $a_B=0.6 a_F$.

\subsection{Attractive Fermi gas in 2D}

Recently Makhalov, Martiyanov, and 
Turlapov (2014) have realized a quasi-2D Fermi
system with widely tunable s-wave interactions
nearly in a ground state, investigating an
ultracold gas of atoms by measuring the pressure $P$ as a function of
the density $n$. The experiment of Makhalov {\it et al.} (2014)
covers physically 
different regimes corresponding to weakly or strongly attractive
Fermi gases or a Bose gas of tightly bound pairs of fermions. 

\begin{figure}[t]
\centerline{\epsfig{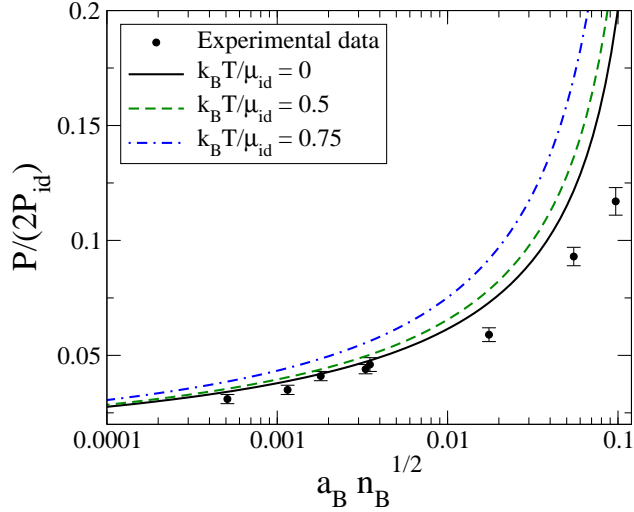}}
\small
\caption{Scaled pressure $P/(2P_{id})$
of the 2D gas of composite bosons as a function of the bosonic 
gas parameter $a_B n_B^{1/2}$, 
where $P_{id}=2\pi\hbar^2 n_B^2/m_B$ is the pressure of an
ideal 2D gas with $m_B$ the mass of each bosonic particle, $a_B$ is the
s-wave scattering length of bosons, and $n_B$ is the bosonic 2D density.
The filled circles with error bars are the experimental data
of Makhalov {\it et al.} (2014).
The curves are obtained from Eqs. (\ref{eb-af}), (\ref{figata})
(\ref{n-mio}) and (\ref{p-mio}) for different values of the scaled
temperature  $k_bT/\mu_{id}$, with $\mu_{id}=4\pi\hbar^2n_B/m_B$. 
Notice that the BCS-BEC mean-field theory  
in the BEC regime predicts a constant pressure, 
independent of the scattering length.}
\label{fig5}
\end{figure}

Within our one-loop Gaussian approach, 
the pressure $P$ is immediately obtained from Eqs. (\ref{malu})
and (\ref{moana}) using the thermodynamic relation $\Omega = - P L^2$:
\beqa
P &=& {m_B \over 8\pi\hbar^2}
\ln{\left({\epsilon_b\over \mu_B } \right)}\, \mu_B^2
\nonumber
\\
&-& {1\over \beta \, L^2}
\sum_{{\bf q}} \ln{\left(1 - e^{-\beta E_{col}(q)} \right)} \; ,  
\eeqa
because, in practice, the gapped 
single-particle fermionic excitations $E_{sp}(k)$ do not contribute to
thermal properties since  $\beta \Delta_0$ is extremely
large in the BEC regime. Morever, the density $n_B$ is given by
\beq
n_B = \left( {\partial P\over \partial \mu_B}\right)_{T,L^2} \; ,
\eeq
from which one finds
\beqa
n_B &=& {m_B \over 4\pi\hbar^2}
\ln{\left({\epsilon_b\over \mu_B \, e^{1/2}} \right)}\, \mu_B  
\nonumber
\\
&-& {1\over \beta\, L^2}
\sum_{{\bf q}} {\partial E_{col}(q)\over \partial \mu_B}
{1\over e^{\beta E_{col}(q)} - 1} \; .
\eeqa
We now use $E_{col}(q)\simeq (\mu_B/m_B)^{1/2} \hbar q$
and the continuum limit $\sum_{\bf q}\to L^2\int d^2{\bf q}/(2\pi)^2$.
In this way we get
\beq
n_B =  {m_B \over 4\pi\hbar^2} \, \mu_B  
\left[ \ln{\left({\epsilon_b\over \mu_B \, e^{1/2}} \right)} - 2
\zeta(3) \left( k_B T\over \mu_B \right)^3 \right] \;  
\label{n-mio}
\eeq
and clearly also
\beq
P =  {m_B \over 8\pi\hbar^2} \, \mu_B^2
\left[ \ln{\left({\epsilon_b\over \mu_B } \right)} + 4 \zeta(3)
\left( k_B T\over \mu_B \right)^3 \right] \; ,
\label{p-mio}
\eeq
where $\zeta(x)$ is the Riemann zeta fuction and $\zeta(3)=1.20205$.
Eqs. (\ref{n-mio}) and (\ref{p-mio}) give, at fixed $k_BT/\mu_B$,
a parametric formula for the the pressure $P$
as a function of the density $n_B$ where $\mu_B$ is the
dummy parameter.

Taking into account Eqs. (\ref{eb-af}) and (\ref{figata}), in Fig. \ref{fig5}
we plot the pressure $P$ in units of the ideal pressure
$P_{id} = 2\pi\hbar^2 n_B^2/m_B$ as a function
of the adimensional gas parameter $a_B n_B^{1/2}$.
In the deep weak-coupling regime $a_B n_B^{1/2} < 0.01$ and
at very low temperature $k_BT/\mu_{id}\ll 1$ with 
$\mu_{id}=4\pi\hbar^2n_B/m_B$, 
the figure shows a good agreement between the experimental data
of Makhalov {\it et al.} (2014) and our theoretical curves. 
Actually the figure suggests 
that the atomic cloud of Makhalov {\it et al.} (2014) was
practically at zero temperature. The deviations between theory and
experiments at larger values of the gas parameter $a_B n_B^{1/2}$ are
presumably due to incomplete bosonization of fermionic pairs. 

It is important to stress that the bosonic collective excitations 
$E_{col}(q)$ are given by Eq. (\ref{ex-bosonic}) only in the 
deep BEC regime. In the full BCS-BEC crossover, 
$E_{col}(q)$ can be obtained numerically setting 
$det({\bf M}(Q))=0$, where ${\bf M}(Q)$
is the inverse propagator for the pair fluctuations, which appears 
in the Gaussian action (see \cite{diener2008})
\beq 
S_{g}[\eta,\bar{\eta},\Delta_0] = {1\over 2} \sum_{Q} 
({\bar\eta}(Q),\eta(-Q)) \ {\bf M}(Q) \left(
\begin{array}{c}
\eta(Q) \\ 
{\bar\eta}(-Q) 
\end{array}
\right) \;
\label{eq:sg-m} 
\eeq
having introduced the Fourier-transformed version of the fluctuation 
fields, with $Q=(\mathrm{i} \Omega_n, \mathbf{q})$, $\Omega_n = 
2 \pi n / \beta$ being the Bose Matsubara frequencies. 
The matrix elements of ${\bf M}(Q)$ are defined by 
\beqa
\mathbb{M}_{11} (Q) = \frac{1}{g} + \sum_\mathbf{k} 
\left( \frac{u^2 u'^2}{\mathrm{i} \omega_n - E - E'} - 
\frac{v^2 v'^2}{\mathrm{i} \omega_n + E + E'} \right) \\
\mathbb{M}_{12} (Q) = \sum_\mathbf{k} u v u' v' 
\left( \frac{1}{\mathrm{i} \omega_n + E + E'} - \frac{1}{\mathrm{i} 
\omega_n - E - E'} \right)
\eeqa
where $u=u_{\mathbf{k}} = \sqrt{\frac{1}{2} ( 1 + 
\frac{\epsilon_\mathbf{k} - \mu}{E_{sp} (\mathbf{k})})}$, 
$v=v_\mathbf{k}=\sqrt{1-u^2_{\mathbf{k}}}$, $u'=u_{\mathbf{k}+
\mathbf{q}}$, $v'=v_{\mathbf{k}+\mathbf{q}}$, $E=E_{sp} (\mathbf{k})$, 
$E'=E_{sp} (\mathbf{k} + \mathbf{q})$. The remaining matrix elements 
are defined by the relations: $ \mathbb{M}_{22} (q)  = 
\mathbb{M}_{11} (-q) $, $ \mathbb{M}_{21} (q)  = \mathbb{M}_{12} (q) $. 

Quite remarkably, \cite{he2015} have numerically found 
a very good agreement between the experimental 
data of Makhalov {\it et al.} 
(2014) and the one-loop Gaussian theory in the full BCS-BEC 
crossover by using the zero-temperature grand potential 
\beq
\Omega_g (\mu,T) = {1\over 2\beta} \sum_{Q} \ln \left[ 
\frac{\mathbb{M}_{11} (Q)}{\mathbb{M}_{22} (Q)} \mbox{det}({\bf M}(Q)) 
\right] e^{\mathrm{i} \Omega_n 0^+} \; , 
\label{eq:omegagreg}
\eeq
which is regularized by convergence factors (\cite{diener2008,he2015}). 
As previously discussed, in the BEC regime of composite bosons 
He {\it et al.} (2015) have recovered our analytical result, 
Eq. (\ref{eb-af}). In a very recent paper \citet{bighin2016}  have compared 
the first sound velocity $c_s$ of the 2D Fermi superfluid 
with preliminar experimental data of  \cite{luick2014}, 
taking into account Eq. (\ref{eq:omegagreg}) and the zero-temperature 
thermodynamic relation 
\beq
c_s = \sqrt{\frac{n}{m} \frac{\partial \mu}{\partial n}} = 
\sqrt{- \frac{n}{m} \left( \frac{1}{L^2} \frac{\partial^2 \Omega (\mu)}
{\partial \mu^2} \right)^{-1}} \; .    
\label{eq:cs}
\eeq
At the mean-field level one has $c_s(\mu_{mf})=v_F/\sqrt{2}$ across the 
whole BCS-BEC crossover, $v_F$ being the Fermi velocity. 
The inclusion of Gaussian fluctuations 
gives a quite different sound velocity: it slowly tends 
to the aforementioned value in the BCS limit, showing, on the 
other hand, a remarkable difference in the intermediate and BEC regimes. 
The preliminar experimental data of \cite{luick2014} are in very good 
agreement with the Gaussian theory of \citet{bighin2016}.  

We can conclude that the one-loop Gaussian theory of fermionic superfluids 
shows good agreement with very recent experimental 
data (\cite{makhalov2014,luick2014}, and also \cite{enss2016}) in the full 
2D BCS-BEC crossover when collective bosonic excitations are appropriately 
taken into account. Moreover, in the BEC regime of the crossover 
the Gaussian theory becomes analytically tractable and, as expected, 
it gives the one-loop equation of state of 2D composite bosons 
whose interaction is characterized by an s-wave  scattering 
length nontrivially related to the scattering length of the atomic fermions. 

\subsection{Attractive Fermi gas in 1D}

The observation of pairing phenomena in a 1D Fermi gas of $^{40}$K atoms 
was reported by \cite{moritz2005}. Using radio-frequency spectroscopy 
they measured the binding energy of two-particle bound 
states of atoms confined in a one-dimensional matter waveguide. 
More recently, \cite{liao2010} measured 
density profiles of 1D trapped two-spin-component fermionic $^6$Li atoms. 
In particular, \cite{liao2010} analyzed the effect of spin imbalance in 
the 1D trapped gas finding a partially polarized 
core surrounded by wings which, 
depending on the degree of polarization, are composed of either a 
completely paired or a fully polarized Fermi gas. This kind of phase 
separation confirms the key features of the phase diagram 
predicted from the exact Bethe-ansatz solution of the 1D unbalanced 
uniform Fermi gas (\cite{orso2007}). As previously stressed, 
for 1D systems (both bosonic and fermionic) the one-loop Gaussian 
theory is fully reliable only in the weak-coupling regime. 

\section{Conclusions}

We have shown that the zero-point energy of both bosonic and fermionic
ultracold atoms contains a finite contribution which plays
a relevant role in the determination of a reliable
equation of state. In the case of repulsive bosonic atoms the final convergent
equation of state, which depends on the dimensionality of the system, 
is independent of the regularization procedure. 
On the contrary, we have found that the dimensional regularization 
cannot be used for three-dimensional attractive fermionic atoms which 
exhibit a BCS-BEC crossover. In fact,  the sign of the scattering length
whose change from negative to positive
across the crossover may be accounted for by using a cutoff 
regularization, would remain always negative under dimensional regularization.
However, the dimensional regularization can be used for the study of
the two-dimensional BCS-BEC crossover because in the two-dimensional
problem the scattering length does not change sign.
By using momentum-cutoff regularization
in the three-dimensional case \citep{sala-giacomo} and
dimensional regularization in the two-dimensional case \citep{sala-flavio}
one derives meaningful equations of state
in the BEC regime of composite bosons. Quite remarkably,
from these equations of state
one obtains simple analytical formulas between the scattering
length of composite molecular bosons and the scattering length
of atomic fermions \citep{sala-giacomo,sala-flavio}. 
Finally, for the one-dimensional Fermi superfluid we have found 
that Gaussian fluctuations improve the mean-field theory 
but do not give the correct equation of state 
in the Tonks-like regime of impenetrable bosons. 

There are several open problems for the physics of ultracold atoms 
which can be faced employing the regularization 
techniques of Gaussian fluctuations discussed in this paper. 
In the two-dimensional case (for both bosonic and fermionic superfluids) 
the Berezinsky-Kosterlitz-Thouless critical 
temperature \citep{berezinskii1971,kosterlitz1973} 
of the superfluid-normal phase transition can be extracted 
by using the Thouless criterion \citep{nagaosa1999} and an accurate description 
of the superfluid density which takes into account Gaussian 
fluctuations in the finite-temperature equation of state. 
Gaussian contributions to the equation of state 
are clearly relevant for Bose-Fermi mixtures \citep{nishida2006}
and for unbalanced superfluid fermions \citep{klimin2012}. For superfluid 
fermionic atoms in three and two dimensions one can also investigate 
the effects of Gaussian fluctuations on the zero-temperature 
condensate fraction \citep{fuku2007} comparing with mean-field 
results \citep{sala-cond3d,sala-cond2d} and available 
Monte Carlo calculations \citep{astra2005}. In conclusion, we stress that 
in addition to ultracold atomic gases, there are several other 
superfluid quantum many-body systems where the methods of functional 
integration and regularization of Gaussian fluctuations 
can play a relevant role to achieve a meaningful and reliable 
theoretical description. Among them we quote 
high-T$_c$ superconductors \citep{scalapino2012}, 
polar molecules in bilayers \citep{zinner2012}, 
neutron matter in the BCS-BEC crossover \citep{sala-nuclear}, 
quark-gluon plasma \citep{bhattacharya2014}, quark matter 
in stars \citep{anglani2014}, exciton-polariton condensates 
\citep{byrnes2014} and, more generally, quantum fluids 
of light \citep{carusotto2013}. 

\section*{Acknowledgments}

The authors acknowledge Ministero Istruzione
Universita Ricerca (PRIN project 2010LLKJBX) for partial support.
The authors thank Lara Benfatto, Giacomo Bighin, Massimo Capone, Milton Cole, 
Luca Dell'Anna, Sergei Klimin, Pieralberto Marchetti, Andrea Perali, 
Pierbiagio Pieri, Carlos Sa de Melo, Adriaan Schakel, 
Giancarlo Strinati, Jacques Tempere, and Andrea Trombettoni 
for enlightening discussions.

\section*{References}

\bibliographystyle{apsrmp}

\end{document}